\documentstyle[12pt,eqsection]{article}
\newcommand{\be}{\begin{equation}}
\newcommand{\ee}{\end{equation}}
\newcommand{\bea}{\begin{eqnarray}}
\newcommand{\eea}{\end{eqnarray}}
\newcommand{\norsl}{\normalsize\sl}
\newcommand{\norsc}{\normalsize\sc}
\def \Lsl {L \kern-.55em{/}}
\def \Ksl {K \kern-.70em{/}}
\def \Psl {P \kern-.65em{/}}
\def \nsl {n \kern-.45em{/}}
\def \Qsl {Q \kern-.65em{/}}
\begin{document}

\begin{titlepage}

\title{ Gauge-independent resummed gluon self-energy  \\ 
in hot QCD }
\author{
\norsc  Ken SASAKI\thanks{e-mail address: sasaki@ed.ynu.ac.jp}\\
\norsl  Dept. of Physics, Yokohama National University\\
\norsl  Yokohama 240, JAPAN\\
\\
\\
\\
In memory of Russell H. Swanson}

\date{}
\maketitle
 
\begin{abstract}
{\normalsize The pinch technique (PT) is applied to obtain 
the gauge-independent resummed gluon self-energy in a hot Yang-Mills gas. 
Calculation is performed at one-loop level in linear gauges 
which preserve rotational invariance and the resummed 
propagators and vertices are used.  The  
effective gluon self-energy, which is obtained as the sum of the resummed gluon 
self-energy and the resummed pinch contributions, 
is not only gauge-independent but also satisfies the transversality 
relation. Using this gauge-independent effective gluon self-energy, we 
calculate the damping rate for transverse gluons in the leading order and show 
that the result coincides with the one obtained by Braaten and Pisarski. 
}
\end{abstract}
 
\begin{picture}(5,2)(-290,-600)
\put(2.3,-50){YNU-HEPTh-96-09-01}
\end{picture}

\thispagestyle{empty}
\end{titlepage}
\setcounter{page}{1}
\baselineskip 18pt

\section{Introduction}
\smallskip

The knowledge of the behaviour of the QCD effective coupling constant 
$\alpha_s (=g^2/{4\pi})$ at high temperature is very important 
for the study of the quark-gluon plasma and/or 
the evolution of the early Universe. The running of $\alpha_s$ with the 
temperature $T$ and the external momentum $k=\vert {\bf k}\vert$ 
is governed by the thermal $\beta$ function $\beta_T$~\cite{Umezawa}. 
However, the previous calculations of $\beta_T$ have exposed various 
problems~\cite{Fujimoto}-\cite{Nakkagawa}, a serious one of which is 
that the results are gauge-fixing dependent~\cite{Niegawa}. 
To circumvent the difficulty concerning the gauge dependence, 
it was then proposed to 
use~\cite{Landsman} the Vilkovisky-DeWitt effective 
action~\cite{Vilkovisky}\cite{Rebhan} or to 
use~\cite{ACPS}-\cite{EK} the background field method (BFM) 
for the calculation of $\beta_T$ at one-loop. 
(In Yang-Mills theories the Vilkovisky-DeWitt effective 
action formalism coincides with BFM in the 
background Landau gauge~\cite{Rebhan}.)

The thermal $\beta$ function $\beta_T$ was calculated in BFM at one-loop level 
for the cases of the gauge parameter 
$\xi_Q=0$~\cite{Landsman}\cite{Eijck}, 
$\xi_Q=1$~\cite{ACPS} and 
$\xi_Q=$an arbitrary number~\cite{EK}. The results are 
expressed in a form, 
\be
    \beta_T^{BFM}=\frac{g^3 N}{2}\biggl\{
              \frac{7}{16}-\frac{1}{8}(1-\xi_{Q})+
                    \frac{1}{64}(1-\xi_{Q})^2 \biggr\}\frac{T}{k},
\label{BetaBFM}
\ee
where $N$ is the number of colors. 
Contrary to the case of the QCD $\beta$ function at zero temperature, 
$\beta_T^{BFM}$ is dependent on the gauge-parameter $\xi_Q$.
This dependence comes from the $\xi_Q$-dependence of the finite part of 
$\Pi_{\mu\nu}$ calculated in BFM. The notion that 
BFM gives $\xi_Q$-dependent finite part for $\Pi_{\mu\nu}$ 
had already been known~\cite{Vilkovisky}\cite{Rebhan}.

Quite recently, it was shown~\cite{Sasakia} 
by the present author that the pinch technique (PT) gives  the 
gauge-independent result for $\beta_T$. 
Calculations  are performed at one-loop level
in four different gauges, 
(i) the background field method with an arbitrary gauge, 
(ii) the Feynman gauge, (iii) the Coulomb gauge, and (iv) 
the temporal axial gauge, and they yield the same result 
$\beta_T=g^3 N \frac{7}{32}\frac{T}{k} $ in all four cases.  
This gauge-independent result for  $\beta_T$ from PT 
corresponds to the result from BFM in Eq.(\ref{BetaBFM}) with the special value 
$\xi_Q=1$ of the gauge parameter.  

However, the above gauge-independent result for  $\beta_T$ 
is incomplete. As Elmfors and Kobes 
pointed out~\cite{EK}, the leading contribution to $\beta_T$, 
which gives a term $T/k$,  does not come from 
the hard part of the loop integral, responsible  for a $T^2/k^2$ term, 
but from soft loop integral. Hence they emphasized that it is not consistent 
to stop the calculation at one-loop order for soft internal momenta and that  
the resummed propagator and vertices~\cite{BPa} must be used to get the 
complete leading contribution. The need for resummation is urged also by 
the following observation: The fact that in 
Ref.\cite{Sasakia}  
the same $\beta_T$ was obtained at one-loop level in four different gauges
implies that the effective gluon self-energy 
$\widehat{\Pi}^{\mu \nu}$, constructed in the framework of    
PT without resummation, is gauge-fixing independent and universal. 
Provided that we use $\widehat{\Pi}^{\mu \nu}$ for calculation of the 
damping rate $\gamma_t$ for  transverse gluons at zero momentum, we would obtain 
$\gamma_t=-\frac{11}{24\pi}Ng^2T$~\cite{HZ}, a negative damping rate, which 
is not acceptable today~\cite{BPb}\cite{KKR}.

The PT is an algorithm to construct, order by order in the coupling 
constant $g$ (in other words, loop by loop), modified {\it gauge-independent} 
off-shell amplitudes through the rearrangement of the Feynman graphs. 
When a theory contains more than one scales like thermal field theories, 
then the contributions from the higher-loop diagrams cannot be neglected 
anymore and the PT fails to give a right answer. On the other hand, 
the resummation in thermal field theories collects systematically 
all leading higher-loop contributions~\cite{Pisa} but 
the effective amplitudes obtained by resummation, in general, contain 
terms which are {\it gauge-dependent}. For the calculation 
of $\beta_T$, we need not only the resummation but also some 
prescription such as the PT to pick up the gauge-independent pieces from the 
effective {\it off-shell} gluon self-energy.

In this paper we will employ the PT and construct the gauge-independent 
resummed gluon self-energy at one-loop order in hot QCD. 
This is the first step toward 
obtaining the gauge-independent thermal $\beta$ 
function $\beta_T$ in the complete leading order. 
Calculation is performed in linear gauges which preserve rotational invariance
and the resummed gluon propagator and vertices are used. 
We find that the resummed effective gluon self-energy, which is the sum of 
the gluon self-energy and the pinch contributions, is 
not only gauge-independent but also satisfies the transversality relation.
Then using this gauge-independent effective gluon self-energy, we  
calculate the damping rate for transverse gluons in the leading order and show 
that the result coincides with one obtained by Braaten and Pisarski~\cite{BPb}.
It should be emphasized that our approach to calculate the gluon damping rate 
is quite different from the one taken by Braaten and Pisarski. In Ref.\cite{BPb} 
they calculated,  in the Coulomb gauge, the effective gluon self-energy 
$^{*}\Pi ^{\mu\nu}$  
using the resummed gluon propagators and vertices. 
The gluon self-energy $^{*}\Pi ^{\mu\nu}$ thus obtained 
contains both gauge-independent and dependent pieces. 
To isolate the gauge-independent pieces in $^{*}\Pi ^{\mu\nu}$ and to 
calculate the gluon damping rate, Braaten and Pisarski constructed the 
two-gluon $\cal T$-matrix element 
by putting $^{*}\Pi ^{\mu\nu}$ on the mass-shell and sandwiching it between 
physical wave functions.  And the gluon damping rate came from the 
imaginary part of the $\cal T$. Our approach is to first 
construct the {\it gauge-independent} effective resummed gluon 
self-energy  $^{*}\widehat{\Pi}^{\mu\nu}$ with recourse to the PT and then to 
calculate the gluon damping rate by putting $^{*}\widehat{\Pi}^{\mu\nu}$ 
on the mass-shell and taking its imaginary part. 
Since $^{*}\widehat{\Pi}^{\mu\nu}$ is already gauge independent, 
we do not have to construct the two-gluon $\cal T$-matrix element.

The PT was proposed some time ago by
Cornwall~\cite{rCa} for an algorithm to 
form new gauge-independent proper vertices and new propagators 
with gauge-independent self-energies. 
First it was used to obtain the one-loop 
gauge-independent effective gluon self-energy and vertices in 
QCD~\cite{rCP}\cite{rPQCD} and then it has been 
applied to the standard model~\cite{StandardModel}. 
Recently the independence of the PT results on the gauge-fixing procedure 
and the uniqueness of the PT algorithm were discussed by 
Papavassiliou and Pilaftsis~\cite{PapaPila}.  Also the PT has been employed to 
show the dual gauge fixing property of the $S$-matrix~\cite{PapaPhil}. 
The application of PT to QCD at high temperature was first made by 
Alexanian and Nair~\cite{Nair} to calculate the gap equation for 
the magnetic mass to one-loop order. The 
issue of gauge independence of several quantities in hot QCD was 
recently discussed from the PT point of view~\cite{Massimo}.

The paper is organized as follows. 
Throughout this paper we work with linear gauges which secure 
rotational symmetry (in the rest frame of the heat bath)~\cite{Kunstattera}. 
So in the next section we present the Feynman rules in this class of gauges.   
Then we review some properties of the resummed two-point function and resummed  
three- and four-point vertices of gluons 
and the Ward-Takahashi identities satisfied by these functions which 
will be in full use in the following sections.
We divide the resummed gluon propagator into four pieces; two pieces are 
gauge independent and the other two are gauge dependent. 
In Sect.3, we consider the resummed gluon self-energy at one loop. 
We decompose it into the sum of several terms according to its gauge dependence.
In Sect.4, we first develop the general prescription necessary 
for extracting the resummed pinch contributions 
to the resummed gluon self-energy from 
the one-loop quark-quark scattering amplitude. Resorting to
this prescription, we then calculate the one-loop resummed pinch contributions. 
In Sect.5, we show that the effective gluon self-energy, which 
is the sum of the resummed gluon self-energy obtained in Sect.3 and 
pinch contributions in Sect.4, is gauge-independent and also 
satisfies the transversality relation. Using the effective gluon self-energy, 
in Sect.6, we calculate the damping rate for the transverse gluons at rest  
in the leading order  and show that 
the result coincides with one obtained by Braaten and Pisarski.
Sect.7 is devoted to the conclusions and discussions.
In addition, we present three Appendices. 
In Appendix A, we give 
one-loop resummed pinch contributions to the gluon self-energy 
from the vertex diagrams of the first kind, of the second kind and 
box diagrams, separately, in the linear gauges which preserve rotational 
invariance. 
In Appendix B, we present a decomposition of the hard thermal loop for 
three-gluon vertex into the sum of terms with different tensor bases.
In Appendix C, we give the expressions of effective three- and four-gluon 
vertices evaluated at ${\bf k=0}$ and $k_0=m_g$,  
where $m_g$ is a thermal gluon mass.  These expressions are necessary 
for the calculation of the gluon damping rate in Sec.6.


\bigskip
\section{Preliminaries}
\smallskip

In this section we introduce all the ``ingredients" which we will 
use in the following sections. 
Throughout this paper we use the Minkowski metric $(+---)$  and  
the notation that 
upper-case letters represent four-momenta:$K^{\mu}=(K^0, {\bf k})$ 
and $k=\vert {\bf k}\vert$. 

For simplicity we consider only gluons. In fact, in Sect. 4 
we introduce quark fields and study the quark-quark scattering at one-loop 
level. However, this is only for picking up the pinch contributions 
of gluons to the resummed gluon self-energy. 
We follow Baier, Kunstatter and Schiff~\cite{Kunstattera} for the choice of gauges. 
We restrict our consideration to linear gauges in the following form:
\be
   I_{\rm gauge~fixing} = -\frac{1}{2\xi}\int d^4 x 
                            \bigl( {\cal F}^a \bigr)^2 ~,   
\ee
with 
\be
      {\cal F}^a = {\cal J}^{\mu} (x) A^a_{\mu}(x)~.
\ee
It may be natural to work in the rest frame of the heat bath where 
rotational symmetry is unbroken. 
The most general gauge fixing condition ${\cal J}^{\mu}$ 
which preserves rotational invariance is 
(in the momentum space representation)
\be
    {\cal J}^{\mu} = c(K) K^{\mu} + b(K) n^{\mu}
\ee
with $n^{\mu}=(1,0,0,0)$.
Now we introduce a four-vector $\widetilde n_{\mu}(K)$ which is the component of 
$n_{\mu}$ orthogonal to the four-momentum $K_{\mu}$~,
\be
   \widetilde n_{\mu}(K) = n_{\mu}- \frac{k_0 K_{\mu}}{K^2}~.
\ee
so that $\widetilde n_{\mu}(K) K^{\mu} =0$~. 
Note 
\be
    \widetilde n^2(K)=\widetilde n_0(K)= -\frac{{\bf k}^2}{K^2}
\ee
In terms of the resulting orthogonal bases, the 
gauge fixing condition ${\cal J}^{\mu}$ is rewritten as 
\be
     {\cal J}^{\mu}=a(K) K^{\mu} + b(K) \widetilde n^{\mu}(K)~,
\ee
where
\be
        a(K)=c(K)+b(K)\frac{k_0}{K^2}~.
\ee

Here we give a few examples of linear gauges: \\
(i) The covariant gauge: 
The gauge fixing condition is given by $\partial^{\mu} A^a_{\mu} = 0$. 
Thus  ${\cal J}^{\mu}_{COV} = K^{\mu}$ and we have 
\be
    a_{COV}=1, \ \ \ b_{COV} =0~.
\ee

\noindent
(ii) The Coulomb gauge : Since $\partial_i A^a_i = 0$ is 
the gauge fixing condition, we have  
${\cal J}^{\mu}_{CG} = K^{\mu}- k_0 n^{\mu}$ and 
\be
    a_{CG}=-\frac{{\bf k}^2}{K^2}=\widetilde n^2(K), \ \ \ b_{CG} =-k_0~.
\ee

\noindent
(iii) The temporal axial gauge: 
The gauge fixing condition is given by $n^{\mu} A^a_{\mu} = 0$. 
Thus  ${\cal J}^{\mu}_{TAG} = -in^{\mu}$ and 
\be
    a_{TAG}=-i\frac{k_0}{K^2}, \ \ \ b_{TAG} =-i~.
\ee
In the temporal axial gauge the gauge parameter 
$\xi$ has a dimension of ${\rm mass}^{-2}$. 

\medskip

For later convenience we introduce two projection operators:
\bea
    P_{\mu \nu}(K)&=&g_{\mu \nu}-\frac{K_{\mu}K_{\nu}}{K^2}
    -\frac{1}{\widetilde n^2(K)}\widetilde n_{\mu}(K) \widetilde n_{\nu}(K) 
       \nonumber \\
     &=& -\pmatrix{0&0\cr
                   0&\delta_{ij}-\frac{k_ik_j}{{\bf k}^2}\cr} \\
  \nonumber \\
    Q_{\mu \nu}(K)&=&
  \frac{1}{\widetilde n^2(K)}\widetilde n_{\mu}(K) \widetilde n_{\nu}(K)~.
\eea
The operator $P_{\mu \nu}(K)$ projects out spatially transverse modes, and 
is orthogonal to both $K^{\mu}$ and $\widetilde n^{\mu}(K)$. 
On the other hand, the operator $Q_{\mu \nu}(K)$ picks up 
longitudinal modes. 
These operators satisfy
\bea
   & & P^2=P~, \ \ \ Q^2=Q~,\ \ \ PQ=QP=0  \\
   & & K^{\mu}P_{\mu \nu}(K)=n^{\mu}P_{\mu \nu}(K)=0  \\
   & & K^{\mu}Q_{\mu \nu}(K)=0~, \ \ \ 
            n^{\mu}Q_{\mu \nu}(K)=\widetilde n_{\nu}(K)  \\
   & & P_{\mu \nu}(K)+Q_{\mu \nu}(K)=g_{\mu \nu}-\frac{K_{\mu}K_{\nu}}{K^2}~.
\eea

\smallskip

In this class of linear gauges, 
the inverse of the bare gluon propagator 
$iD^{(0)ab}_{\mu \nu}=i\delta ^{ab}D^{(0)}_{\mu \nu}$ is given by 
\be
   D^{(0)-1}_{\mu \nu}(K)=\Gamma_{\mu \nu}(K)-
       \frac{1}{\xi}(a K_{\mu}+b \widetilde n_{\mu})
                    (a K_{\nu}+b \widetilde n_{\nu})
\ee
where
\bea
    \Gamma_{\mu \nu}(K)&=&-K^2 g_{\mu \nu} + K_{\mu}K_{\nu} \nonumber \\
              &=&-K^2 \bigl[P_{\mu \nu}(K)+Q_{\mu \nu}(K)\bigr]   
\eea
is the bare two-point function.
The ghost propagator is 
\be
    iG^{ab}(K)=i\delta ^{ab} \frac{-1}{a(K)K^2} 
\ee
and the ghost-gluon vertex (see Fig.1) is expressed as 
\be
   gf^{abc}\Gamma_{\mu}=gf^{abc}
         \bigl[ a(P)P_{\mu}+b(P)\widetilde n_{\mu}(P) \bigr] ~.
\ee 

\smallskip

The bare gluon three- and four-point vertices depend only on the classical action 
owning to the linearity of the gauge fixing condition.  
Thus the bare three-point vertex  
$\Gamma_{\mu \nu \lambda}^{abc} (P, Q, R) \equiv 
gf^{abc}\Gamma_{\mu \nu \lambda} (P, Q, R)$ 
and the bare four-point vertex  
$\delta^{cd} \Gamma_{\mu \nu \alpha \beta}^{abcd} (P, Q, R, S) \equiv 
-i\delta^{ab}g^2 N\Gamma_{\mu \nu \alpha \beta} (P, Q, R, S)$ are, respectively, 
expressed as 
\bea
   \Gamma_{\mu \nu \lambda} (P, Q, R)&=&(P-Q)_{\lambda}g_{\mu \nu}
      +(Q-R)_{\mu}g_{\nu \lambda}+(R-P)_{\nu}g_{\lambda \mu}  \\
  \Gamma_{\mu \nu \alpha \beta} (P, Q, R, S)&=&
    2g_{\mu \nu }g_{\alpha \beta}-g_{\mu \alpha}g_{\nu \beta}
                  -g_{\mu \beta}g_{\nu \alpha}
\eea
where it is understood that  
each momentum flows inward and thus the relations  $P+Q+R=0$ 
and $P+Q+R+S=0$ hold, respectively, in the three- and four-point vertices. 
Also we have traced over the last two color indices in the four-point vertex. 
These bare two-point function and three- and four-point vertices satisfy the 
following Ward-Takahashi identities:
\bea
    & & K^{\mu}\Gamma_{\mu \nu}(K)=0  \\
    & & R^{\lambda}\Gamma_{\mu \nu \lambda} (P, Q, R) = 
             \Gamma_{\mu \nu}(P) - \Gamma_{\mu \nu}(Q)  \\
    & & S^{\beta}\Gamma_{\mu \nu \alpha \beta} (P, Q, R, S) = 
          \Gamma_{\mu \nu \alpha} (P+S, Q, R) - 
                   \Gamma_{\mu \nu \alpha} (P, Q+S, R)    
\eea

In gauge theories at finite temperature we lose 
the usual connection between the loop expansion and the powers of $g$.
In hot QCD there appear two independent mass scales, $T$ and $gT \ll T$. Momenta 
which are of order $T$ are called ``hard", while those which are of order $gT$ 
are called ``soft". If all of the external legs in a bare amplitude are soft, then 
one-loop corrections to that amplitude, in which all internal momenta 
are hard, are of the same order in $g$. In order to calculate consistently, 
we must take into account these one-loop corrections called 
the hard thermal loops. The higher-order effects coming from the hard thermal loops 
are systematically resummed into effective propagators and vertices~\cite{BPa}.

Let $\delta \Pi_{\mu \nu}$, $\delta \Gamma_{\mu \nu \lambda}$ and 
$\delta \Gamma_{\mu \nu \alpha \beta}$ denote the hard thermal 
loop corrections to the two-point function and 
three- and four-point vertices, respectively. 
It was found~\cite{BPa}\cite{Pisa}\cite{Taylor} 
that these hard thermal loops are gauge independent and obey the 
same Ward-Takahashi identities as the tree-level functions.
Then if the corrected two-point function and 
three- and four-point vertices
are defined as 
\bea
    ^{*}\Gamma_{\mu \nu}&=&\Gamma_{\mu \nu}+\delta \Pi_{\mu \nu} \\
    ^{*}\Gamma_{\mu \nu \lambda}&=&\Gamma_{\mu \nu \lambda}+
                    \delta \Gamma_{\mu \nu \lambda}   \\
    ^{*}\Gamma_{\mu \nu \alpha \beta}&=&\Gamma_{\mu \nu \alpha \beta}+
                  \delta  \Gamma_{\mu \nu \alpha \beta} ~,   
\eea
it is obvious that these corrected functions also 
satisfy the same Ward-Takahashi identities. More precisely, we have  
\bea
    & & K^{\mu} {^{*}\Gamma_{\mu \nu}(K)}=0  
\label{WTTwo}   \\
    & & R^{\lambda} {^{*}\Gamma_{\mu \nu \lambda} (P, Q, R)} = 
             {^{*}\Gamma_{\mu \nu}(P)} - {^{*}\Gamma_{\mu \nu}(Q)}  
\label{WTThree}    \\
    & & S^{\beta} {^{*}\Gamma_{\mu \nu \alpha \beta} (P, Q, R, S)} = 
          {^{*}\Gamma_{\mu \nu \alpha} (P+S, Q, R)} - 
                   {^{*}\Gamma_{\mu \nu \alpha} (P, Q+S, R)}~.  
\label{WTFour}  
\eea

Since the hard thermal loop $\delta \Pi_{\mu \nu}$ satisfies the identity
$K^{\mu}\delta \Pi_{\mu \nu}(K)=0$, it can be decomposed as 
\be
  \delta \Pi_{\mu \nu}(K)=\delta \Pi_T (K) P_{\mu \nu}(K) 
          + \delta \Pi_L (K) Q_{\mu \nu}(K) ~.
\ee
Thus ${^{*}\Gamma_{\mu \nu}}$ can be written as 
\be
    {^{*}\Gamma_{\mu \nu}}(K) = -K^2_T P_{\mu \nu}(K) - K^2_L Q_{\mu \nu}(K)~,  
\label{Gamma2}  
\ee
where 
\bea
    K^2_T &=& K^2 - \delta \Pi_T (K)  \\
    K^2_L &=& K^2 - \delta \Pi_L (K) 
\eea
and it satisfies
\bea
   & & n^{\mu}{^{*}\Gamma_{\mu \nu}}(K) =
{^{*}\Gamma_{0 \nu}}(K) = - K^2_L \widetilde n_{\nu} (K) \\
   & & {^{*}\Gamma_{\mu \lambda}}(K) {^{*}\Gamma^{\lambda}}_{\nu}(K) = 
             K^4_T P_{\mu \nu}(K) + K^4_L Q_{\mu \nu}(K)~.
\eea
The explicit forms of $\delta \Pi_T (K)$ and $\delta \Pi_L (K)$ are given in 
Appendix B.

The resummed three-point vertex ${^{*}\Gamma_{\mu \nu \lambda}}$ 
has the same properties as the bare one $\Gamma_{\mu \nu \lambda}$~\cite{BPa}:
\bea
  & & {^{*}\Gamma_{\mu \nu \lambda} (P, Q, R)} = 
      {^{*}\Gamma_{\nu \lambda \mu} (Q, R, P)} =
      {^{*}\Gamma_{\lambda \mu \nu} (R, P, Q)}    \nonumber 
\label{PropThree}         \\
  & & {^{*}\Gamma_{\mu \nu \lambda} (-P, -Q, -R)} = 
                  - {^{*}\Gamma_{\mu \nu \lambda} (P, Q, R)}   \\
  & & {^{*}\Gamma_{\mu \nu \lambda} (P, Q, R)} =
               - {^{*}\Gamma_{\mu \lambda \nu} (P, R, Q)}  \nonumber 
\eea
Also the resummed four-point vertex  
${^{*}\Gamma_{\mu \nu \alpha \beta} (P, Q, R, S)}$ is, just like the bare 
one, symmetric under interchange of momenta and Lorentz indices of the 
first two lines, the last two lines, and interchange of the first pair 
with the second pair. The Ward-Takahashi identities 
Eqs.(\ref{WTTwo})-(\ref{WTFour}) satisfied by 
the corrected two-point function and three- and four-point vertices
and the properties of the corrected three-point vertex given in Eq.(\ref{PropThree}) 
will be frequently used in the following sections to show the 
gauge-independence of the resummed gluon self-energy obtained 
by the PT.

\smallskip

The inverse of the resummed propagator 
$i~{^{*}D}^{ab}_{\mu \nu}=i\delta ^{ab}~{^{*}D}_{\mu \nu}$ is given by 
\bea
    {^{*}D^{-1}_{\mu \nu}}(K) &=& D^{(0)-1}_{\mu \nu}(K) + 
                   \delta \Pi_{\mu \nu}(K)  \nonumber \\
         &=& -K^2_T P_{\mu \nu}(K) - K^2_L Q_{\mu \nu}(K) -
              \frac{1}{\xi}(a K_{\mu}+b \widetilde n_{\mu})
                    (a K_{\nu}+b \widetilde n_{\nu}) ~.
\eea 
Inverting ${^{*}D}^{-1}_{\mu \nu}$, we have for the resummed propagator
\bea
   {^{*}D}_{\mu \nu}(K) &=& -\frac{1}{K^2_T}P_{\mu \nu}(K) - 
     \frac{1}{K^2_L} \biggl[Q_{\mu \nu}(K) -\frac{b}{a}\frac{1}{K^2}
            \bigl(\widetilde n_{\mu}K_{\nu}+K_{\mu}\widetilde n_{\nu}\bigr) +
       \frac{b^2}{a^2}\widetilde n^2 \frac{K_{\mu}K_{\nu}}{K^4} 
        \biggr]    \nonumber \\
   & & - \frac{\xi}{a^2} \frac{K_{\mu}K_{\nu}}{K^4}~. 
\eea
For  later convenience, we rewrite ${^{*}D}_{\mu \nu}$ as follows:
\bea
  {^{*}D}_{\mu \nu}(K) &=& A(K) g_{\mu \nu} + B(K) \frac{K_{\mu}K_{\nu}}{K^2} 
    + S(K)\biggl[n_{\mu}n_{\nu}-\frac{k_0}{K^2}(n_{\mu}K_{\nu}+K_{\mu}n_{\nu})
      \biggr]  \nonumber  \\
     & &+T(K) (n_{\mu}K_{\nu}+K_{\mu}n_{\nu}) 
\label{Propagator} 
\eea
where
\bea
      A(K) &=& - \frac{1}{K^2_T}     
\label{AAA}   \\
      B(K) &=& - \frac{1}{K^2_T}\frac{K^2}{{\bf k}^2} + 
                \frac{1}{K^2_L}\biggl\{ \frac{k^2_0}{{\bf k}^2} 
   -\frac{2k_0}{K^2}\frac{b}{a} + \frac{{\bf k}^2}{K^4}\frac{b^2}{a^2} \biggr\}
       - \frac{\xi}{K^2a^2}  \\
     S(K) &=& \biggl(\frac{1}{K^2_T} - \frac{1}{K^2_L} \biggr) 
                \frac{1}{\widetilde n^2(K)}  
\label{SSS} \\ 
 T(K) &=&   \frac{1}{K^2_L} \frac{1}{K^2} \frac{b}{a}  
\label{TTT}
\eea
Since the ratio $b(K)/a(K)$ is a odd function of $K$, we easily see that 
$A(K)$, $B(K)$, and $S(K)$ are even functions of $K$, while $T(K)$ is an 
odd function. 
Also $A(K)$ and $S(K)$ do not depend on $a$, $b$, nor the gauge parameter $\xi$ 
and thus they are gauge-independent. 
It is noted that the gauge parameter $\xi$ only appears in $B(K)$. 
If we take the limit $\delta \Pi_T=\delta \Pi_L=0$, the function $S(K)$ vanishes.
So the $S(K)$ term is a unique one for the resummed gluon propagator.
In the next two sections we use the decomposed form of 
the gluon propagator given in Eq.(\ref{Propagator}) and divide 
the one-loop resummed gluon self-energy and 
the corresponding resummed pinch contributions into terms according to its 
dependence on the functions $A$, $B$, $S$, and $T$. In doing so, we can 
easily see the cancellation of the gauge-dependent parts when 
the pinch contributions are added to the resummed gluon self-energy.


\bigskip
\section{Resummed Gluon Self-Energy}
\smallskip


In this section 
we consider the resummed gluon self-energy ${^{*}\Pi_{\mu\nu}}$. 
We assume that the external and the loop momenta are soft. Then 
there are three diagrams which contribute at one-loop:
\be
  {^{*}\Pi_{\mu\nu}}(K) ={^{*}\Pi^{3g}_{\mu\nu}}(K)+
         {^{*}\Pi^{4g}_{\mu\nu}}(K)+{\Pi^{gh}_{\mu\nu}}(K)~.
\label{PiSum}
\ee
The graph in Fig.2(a) with two resummed three-gluon vertices gives 
\be
  {^{*}\Pi^{3g}_{\mu\nu}}(K)=\frac{Ng^2}{2} \int dP~ 
      {^{*}\Gamma_{\lambda\mu\alpha}}(P,K,Q)~ {^{*}D^{\alpha\beta}}(Q)~ 
      {^{*}\Gamma_{\beta\nu\tau}}(-Q,-K,-P)~ {^{*}D^{\tau\lambda}}(P)~,
\label{Pi3g}
\ee
with
\be
    \int dP = \int \frac{d^3p}{8\pi^3} T \sum_n~, 
\ee
where the summation goes over the integer $n$ in $p_0 =i 2\pi n T$ and 
the spatial integration is implicitly assumed to be over soft momenta only. 
We have chosen the variables as $K+P+Q=0$ so that there holds a 
relation
\be
  \int dP f(P,Q)=\int dP f(Q,P)~.  
\ee
In the following we make extensive use of this symmetry property of 
the integrands under interchange of $P$ and $Q$. The graph in Fig.2(b) 
with a resummed four-gluon vertex gives  
\be
  {^{*}\Pi^{4g}_{\mu\nu}}(K)=\frac{Ng^2}{2} \int dP~ 
      {^{*}\Gamma_{\mu\nu\alpha\beta}}(K,-K,P,-P)~ {^{*}D^{\alpha\beta}}(P)~. 
\ee
Finally, the contribution of the ghost loop in Fig.2(c) gives 
\bea
  {\Pi^{gh}_{\mu\nu}}(K)&=& Ng^2 \int dP
     \bigl[ a(P)P_{\mu}+b(P)\widetilde n_{\mu}(P) \bigr]
     \bigl[ a(Q)Q_{\nu}+b(Q)\widetilde n_{\nu}(Q) \bigr] \nonumber \\
  & & \qquad  \qquad  \qquad  
             \times \frac{1}{a(P)P^2a(Q)Q^2}  \nonumber \\
  &=& Ng^2 \int dP \frac{1}{P^2 Q^2}
      \biggl\{ P_{\mu}Q_{\nu} + \frac{b(P)}{a(P)}
        \bigl[ \widetilde n_{\mu}(P) Q_{\nu} + 
                 Q_{\mu}\widetilde n_{\nu}(P) \bigr] \nonumber \\
 & & \qquad  \qquad  \qquad  \qquad  \qquad  \qquad +
    \frac{b(P)b(Q)}{a(P)a(Q)}\widetilde n_{\mu}(P) \widetilde n_{\nu}(Q)
    \biggr\}~. 
\label{PiGhost}
\eea

To study the gauge-dependence of ${^{*}\Pi^{3g}_{\mu\nu}}$, 
we use the decomposed form of the resummed propagator 
in Eq.(\ref{Propagator}) and rewrite ${^{*}\Pi^{3g}_{\mu\nu}}$ 
in terms of $A$, $B$, $S$, and $T$ given in Eqs.(\ref{AAA})-(\ref{TTT}).
By virtue of the Ward-Takahashi identities 
satisfied by the resummed vertices, we find that ${^{*}\Pi^{3g}_{\mu\nu}}$ 
is expressed as 
\be
    {^{*}\Pi^{3g}_{\mu\nu}}(K)=Ng^2 \int dP  
            \sum_{i\ge j} I^{ij}_{\mu\nu} \qquad \qquad {\rm with} \quad 
         i, j=A, B, S, T ~,  
\ee 
where
\bea
 & & I^{AA}_{\mu\nu}=-\frac{1}{2} A(P)A(Q)~
            {^{*}\Gamma_{\lambda\mu}}^{\alpha}(P,K,Q)~ 
            {^{*}\Gamma_{\alpha\nu}}^{\lambda}(Q,K,P)~,  
\label{IAA}  \\
 & & I^{BB}_{\mu\nu}=\frac{1}{2} B(P)B(Q)\frac{1}{P^2 Q^2}
           P^{\lambda}P^{\tau}~{^{*}\Gamma_{\mu\lambda}}(K)~
                 {^{*}\Gamma_{\nu\tau}}(K)~,  
\label{IBB}  \\
 & & I^{SS}_{\mu\nu}=-S(P)S(Q)\biggl[~ \frac{1}{2}~
           {^{*}\Gamma_{0 \mu 0}}(P,K,Q)~{^{*}\Gamma_{0 \nu 0}}(Q,K,P) 
   \nonumber   \\
  & & \qquad  \qquad + 
   \frac{p_0}{P^2} \biggl\{ \bigl[K^2_L \widetilde n_{\mu}(K)-
                           Q^2_L \widetilde n_{\mu}(Q) \bigr]~
       {^{*}\Gamma_{0 \nu 0}}(Q,K,P) +
     (\mu \leftrightarrow \nu)  \biggr\}  \nonumber  \\
  & & \qquad  \qquad +\frac{p_0q_0}{P^2Q^2}\biggl\{  
  \bigl[K^2_L \widetilde n_{\mu}(K)- Q^2_L \widetilde n_{\mu}(Q) \bigr]
  \bigl[K^2_L \widetilde n_{\nu}(K)- P^2_L \widetilde n_{\nu}(P) \bigr] 
  \nonumber  \\
  & & \qquad  \qquad \qquad  \qquad 
   -\frac{1}{2}\bigl[ P^{\lambda}~{^{*}\Gamma_{\lambda\mu}}(K)~
   {^{*}\Gamma_{0 \nu 0}}(Q,K,P) + (\mu \leftrightarrow \nu)  \bigr] \biggr\}
   \biggr], 
\label{ISS} \\
 & & I^{TT}_{\mu\nu}=T(P)T(Q)
       \biggl[-\bigl[K^2_L \widetilde n_{\mu}(K)-
                           P^2_L \widetilde n_{\mu}(P) \bigr]
          \bigl[K^2_L \widetilde n_{\nu}(K)-
                           Q^2_L \widetilde n_{\nu}(Q) \bigr]  \nonumber \\
 & & \qquad \qquad \qquad  \qquad +\frac{1}{2} \Bigl\{
       P^{\lambda}~{^{*}\Gamma_{\lambda\mu}}(K)~ 
         {^{*}\Gamma_{0 \nu 0}}(Q,K,P) + (\mu \leftrightarrow \nu) \Bigr\} 
              \biggr]~,  \\
 & & I^{AB}_{\mu\nu}=A(Q)B(P)\frac{1}{P^2}
        \bigl[{^{*}\Gamma_{\mu\alpha}}(K)-{^{*}\Gamma_{\mu\alpha}}(Q) \bigr]
   \bigl[{^{*}\Gamma^{\alpha}}_{\nu}(K)-{^{*}\Gamma^{\alpha}}_{\nu}(Q) \bigr]~, 
\\
 & & I^{AS}_{\mu\nu}=-A(Q)S(P)\biggl[{^{*}\Gamma_{0 \mu \alpha }}(P,K,Q)~
              {^{*}\Gamma^{\alpha}}_{\nu 0}(Q,K,P) \nonumber  \\
  & & \qquad \qquad \qquad - \frac{p_0}{P^2} \Bigl\{  
 \bigl[{^{*}\Gamma_{\mu\alpha}}(K)-{^{*}\Gamma_{\mu\alpha}}(Q) \bigr]~
        {^{*}\Gamma^{\alpha}}_{\nu 0}(Q,K,P) + (\mu \leftrightarrow \nu) 
              \Bigr\} \biggr] 
\label{IAS} \\
 & & I^{AT}_{\mu\nu}=-A(Q)T(P) 
    \biggl\{ \bigl[{^{*}\Gamma_{\mu\alpha}}(K)-
         {^{*}\Gamma_{\mu\alpha}}(Q) \bigr]{^{*}\Gamma^{\alpha}}_{\nu 0}(Q,K,P) 
         + (\mu \leftrightarrow \nu)  \biggr\},  \\
 & & I^{BS}_{\mu\nu}=B(P)S(Q)\frac{1}{P^2} \biggl[
   \bigl[K^2_L \widetilde n_{\mu}(K) - Q^2_L \widetilde n_{\mu}(Q) \bigr]
 \bigl[K^2_L \widetilde n_{\nu}(K) - Q^2_L \widetilde n_{\nu}(Q) \bigr] 
   \nonumber  \\
 & & \qquad \qquad \qquad + \frac{q_0}{Q^2} \Bigl\{ 
  \bigl[K^2_L \widetilde n_{\mu}(K) - Q^2_L \widetilde n_{\mu}(Q) \bigr]~
      Q^{\lambda}~{^{*}\Gamma_{\lambda\nu}}(K) 
 + (\mu \leftrightarrow \nu) \Bigr\} \biggr] \\
 & & I^{BT}_{\mu\nu}=-B(P)T(Q)\frac{1}{P^2}
         \biggl[ Q^{\lambda}~{^{*}\Gamma_{\lambda\mu}}(K) 
   \bigl[K^2_L \widetilde n_{\nu}(K) - Q^2_L \widetilde n_{\nu}(Q) \bigr] 
          + (\mu \leftrightarrow \nu) \biggr] \\
 & & I^{ST}_{\mu\nu}=S(Q)T(P) 
         \biggl[ \Bigl\{ \bigl[K^2_L \widetilde n_{\mu}(K) - 
                     Q^2_L \widetilde n_{\mu}(Q) \bigr]~ 
   {^{*}\Gamma_{0 \nu 0}}(Q,K,P)  + (\mu \leftrightarrow \nu) \Bigr\} 
                \nonumber \\
    & & \qquad \qquad + \frac{q_0}{Q^2} \Bigl\{ 
  \bigl[K^2_L \widetilde n_{\mu}(K) - 
                     P^2_L \widetilde n_{\mu}(P) \bigr]
  \bigl[K^2_L \widetilde n_{\nu}(K) - 
                     Q^2_L \widetilde n_{\nu}(Q) \bigr]
  + (\mu \leftrightarrow \nu) \Bigr\}  \nonumber  \\
  & & \qquad \qquad + \frac{q_0}{Q^2} \Bigl\{ 
     Q^{\lambda}~{^{*}\Gamma_{\lambda\mu}}(K)~{^{*}\Gamma_{0 \nu 0}}(Q,K,P) 
   + (\mu \leftrightarrow \nu) \Bigr\} 
       \biggr]~.  
\label{IST}
\eea

For an illustration, let us show the derivation of $I^{BB}_{\mu\nu}$ term. 
The product of two propagators 
${^{*}D^{\alpha\beta}}(Q)~{^{*}D^{\tau\lambda}}(P)$ in Eq.(\ref{Pi3g}) has a term
\be
          B(P)B(Q)\frac{P^{\lambda}P^{\tau}Q^{\alpha}Q^{\beta}}{P^2 Q^2}~, 
\ee
which gives
\be
    I^{BB}_{\mu\nu}=\frac{1}{2}B(P)B(Q)
             \frac{P^{\lambda}P^{\tau}Q^{\alpha}Q^{\beta}}{P^2 Q^2}~
                   {^{*}\Gamma_{\lambda\mu\alpha}}(P,K,Q)~
        {^{*}\Gamma_{\beta\nu\tau}}(-Q,-K,-P)~.
\ee
Using the properties of the corrected three-point vertex  
${^{*}\Gamma_{\mu \nu \lambda}}$ in Eq.(\ref{PropThree}) and the 
Ward-Takahashi identities (\ref{WTTwo}) and (\ref{WTThree}) satisfied by 
${^{*}\Gamma_{\mu \nu}}$ and ${^{*}\Gamma_{\mu \nu \lambda}}$, we find that  
\be
    P^{\lambda}P^{\tau}Q^{\alpha}Q^{\beta}
    {^{*}\Gamma_{\lambda\mu\alpha}}(P,K,Q)~
        {^{*}\Gamma_{\beta\nu\tau}}(-Q,-K,-P) = 
           P^{\lambda}P^{\tau}~{^{*}\Gamma_{\lambda \mu}}(K)
                 ~{^{*}\Gamma_{\nu \tau}}(K)
\ee 
and thus we reach the expression for $I^{BB}_{\mu\nu}$ given in Eq.(\ref{IBB}).
The other terms in Eq.(\ref{IAA}) through Eq.(\ref{IST}) 
are derived in a similar way.

Likewise ${^{*}\Pi^{4g}_{\mu\nu}}$ is expressed 
in terms of $A$, $B$, $S$, and $T$ as 
\be
    {^{*}\Pi^{4g}_{\mu\nu}}(K)=Ng^2 \int dP  
            \sum_{i} J^{i}_{\mu\nu} \qquad \qquad {\rm with} \quad 
         i = A, B, S, T  ~, 
\ee 
where
\bea 
 & & J^{A}_{\mu\nu}=\frac{1}{2}A(P)~
        {^{*}\Gamma_{\mu\nu\alpha}}^{\alpha}(K,-K,P,-P)  ~, 
\label{JA} \\
 & & J^{B}_{\mu\nu}=B(P)\frac{1}{P^2}
         \bigl[{^{*}\Gamma_{\mu\nu}}(K)-{^{*}\Gamma_{\mu\nu}}(Q) \bigr]~,  \\
 & & J^{S}_{\mu\nu}= S(P)~\biggl[ \frac{1}{2}
     {^{*}\Gamma_{\mu\nu 0 0}}(K,-K,P,-P) \nonumber  \\
  & & \qquad \qquad \qquad \qquad + \frac{p_0}{P^2} \Bigl\{ 
  {^{*}\Gamma_{\mu\nu 0}}(Q,K,P)
                + (\mu \leftrightarrow \nu) \Bigr\} \biggr]~, 
\label{JS} \\
 & & J^{T}_{\mu\nu}= - T(P)
  \Bigl\{{^{*}\Gamma_{\mu\nu 0}}(Q,K,P)
                + (\mu \leftrightarrow \nu) \Bigr\}~.
\eea


\bigskip
\section{Resummed Pinch Contributions}
\smallskip

\subsection{\it Pinch Technique}
\smallskip

In this section we obtain the one-loop resummed pinch contributions 
to the resummed gluon self-energy. We proceed in the same way as we did 
before in the second paper of Ref.\cite{Sasakia}.
The only difference is that here we use the resummed gluon propagators and 
resummed vertices instead of bare ones.
Let us consider the $S$-matrix 
element $\widehat T$ for the elastic quark-quark scattering at one-loop, 
assuming that quarks 
have the same mass $m$. We introduce quarks just as technical devices   
to extract the pinch contributions. We assume that both the momentum 
transfered in the $t$-channel and the loop momenta are soft.
Besides the self-energy diagram in Fig.3, 
the vertex diagrams of the first and second kind   
and the box diagrams contribute to $\widehat T$.
They are shown in Fig.4(a), Fig.5(a), and Fig.6(a), respectively. 
Note that gluon propagators and vertices are resummed ones. 
For quark sectors, however, we still use bare quark propagators and 
bare quark-gluon vertices, because we are only interested in gluonic 
parts. Although each contribution from diagrams in 
Fig.3, Fig.4(a), Fig.5(a), and Fig.6(a)
is, in general, gauge-dependent, the sum 
is gauge-independent. This can be seen from the following observation:
When we set $T=0$, all the hard thermal loop contributions vanish. 
Then the sum reduces to the ordinary zero-temperature $S$-matrix 
element for the elastic quark-quark scattering at one-loop, which is 
obviously gauge-independent. At finite temperature the hard thermal loop 
contributions are switched on. Since these contributions do not depend on the gauge 
choices and thus the sum remains gauge-independent.

Now we single out the ``pinch parts'' of the vertex and box diagrams, 
which are depicted in Fig.4(b), Fig.5(b), and Fig.6(b). 
They emerge when a $\gamma^{\mu}$ matrix on the quark line is 
contracted with a four-momentum $K_{\mu}$ offered by a resummed gluon propagator 
or a  resummed three-gluon vertex. Such a term triggers an elementary Ward 
identity of the form
\be
      \Ksl = ({\Psl} + \Ksl -m) - ({\Psl} -m).
\label{Ward}
\ee
The first term removes (pinches out) the internal quark propagator, 
whereas the second term vanishes on shell, or {\it vice versa} . This 
procedure leads to contributions to $\widehat T$ with one or two less 
quark propagators and, 
hence, we will call these contributions $\widehat T_P$,  
``pinch parts'' of $\widehat T$.

Next we extract from $\widehat T_P$ the pinch contributions to the 
resummed gluon self-energy  
${{^*}\Pi}_{\mu \nu}$. 
First note that the contribution of the resummed gluon self-energy diagram to 
$\widehat T$ is written in the form (see Fig.3)
\be
  \widehat T^{(S.E)}=[T^a \gamma _{\alpha}]~{^{*}D}^{\alpha \mu}(K)
       ~{{^*}\Pi}_{\mu\nu}~{^{*}D}^{\nu \beta}(K)[T^a \gamma _{\beta}],
\ee
where $T^a$ is a representation matrix  
of $SU(N)$, and $\gamma _{\alpha}$ and $\gamma _{\beta}$ are 
$\gamma$ matrices on the external quark lines. 
The pinch contribution ${{^*}\Pi}^P_{\mu \nu}$ to 
$\widehat T_P$ should have the same form. 
Thus we must take away $[T^a \gamma _{\alpha}]{^{*}D}^{\alpha \mu}(K)$ and 
${^{*}D}^{\nu \beta}(K)[T^a \gamma _{\beta}]$ from $\widehat T_P$.  
For that purpose we use the following identity satisfied by 
the resummed gluon propagator and its inverse:
\bea
   {g_{\alpha}}^{\beta}&=&{^{*}D}_{\alpha \mu}(K)[{^{*}D}^{-1}]^{\mu \beta}(K)
      ={^{*}D}_{\alpha \mu}(K)~{^{*}\Gamma^{\mu \beta}}(K) + K_{\alpha}~  {\rm term} 
       \nonumber \\
           &=&{^{*}D}^{-1}_{\alpha \mu }(K)~{^{*}D}^{\mu \beta}(K) =
     {^{*}\Gamma_{\alpha \mu}}(K)~{^{*}D}^{\mu \beta}(K) + K_{\beta}~ {\rm term},
\label{Identity}
\eea
where ${^{*}\Gamma_{\mu \nu}}(K)$ is given in Eq.(\ref{Gamma2}).
The $K_{\alpha}$ and  $K_{\beta}$ terms give  null results  when 
they are contracted with $\gamma_{\alpha}$ and $\gamma_{\beta}$,
respectively,  of the external quark lines.

The pinch part of the one-loop vertex diagrams of the first kind depicted 
in Fig.4(b) plus their mirror graphs has a form 
\be
  \widehat T_P^{(V_1)}={\cal A}[T^a \gamma _{\alpha}]~{^{*}D}^{\alpha \beta}(K)
      [T^a \gamma _{\beta}]~,
\ee
where ${\cal A}$ (also ${\cal B}^0$, ${\cal B}^{1}_{\lambda\nu}$, 
${\cal B}^{2}_{i\nu}$, ${\cal C}^0$, and 
${\cal C}_{ij}$ in the equations below) contains a loop integral. 
Using Eq.(\ref{Identity}) we find 
\be
    \gamma_{\alpha}~{^{*}D}^{\alpha \beta}(K)\gamma _{\beta}=
       \gamma_{\alpha}~{^{*}D}^{\alpha \mu}(K)~{^{*}\Gamma_{\mu \nu}}(K)
     ~{^{*}D}^{\nu \beta}(K)\gamma_{\beta}.
\label{gammagamma}
\ee
Thus  the contributions to  ${^{*}\Pi}_{\mu \nu}$ from the vertex diagrams of 
the first kind are written as 
\be
   {^{*}\Pi}_{\mu \nu}^{P(V_1)}={^{*}\Gamma_{\mu \nu}}(K){\cal A}.
\ee

The pinch part of the one-loop vertex diagrams of the second kind depicted 
in Fig.5(b) has a form 
\be
  \widehat T_P^{(V_2)}=[T^a] \biggl\{ [\gamma_{\nu}] {\cal B}^0 + 
           [\gamma^{\lambda}] {\cal B}^1_{\lambda \nu}+
     \sum_{i}[\Psl_i]{\cal B}^2_{i\nu}   \biggr\} 
   ~{^{*}D}^{\nu \beta}(K) [T^a \gamma _{\beta}]~,
\label{TPV2}
\ee    
where $P_i$ is a four-momenta appearing 
in the diagrams. By redefinition of the loop-integral momentum
we can choose $P_i=P\  {\rm or}\  n$   where $P$ is the 
loop-integral momentum and $n$ is a unit vector $n^{\mu}=(1,0,0,0)$.  
Using Eq.(\ref{gammagamma}) and 
\be
   [\gamma^{\lambda}]  =
  [\gamma_{\alpha}]~{^{*}D}^{\alpha \mu}(K)~{{^{*}\Gamma}_{\mu}}^{\lambda}(K)   
\ee
\be
  [{\Psl}_i] =
   [\gamma_{\alpha}]~{^{*}D}^{\alpha \mu}(K)~{^{*}\Gamma}_{\mu\lambda}(K) 
      P_i^{\lambda},
\ee
we obtain  for the contributions to  ${^{*}\Pi}_{\mu \nu}$ from the vertex 
diagrams of the second kind
\bea
   {^{*}\Pi}_{\mu \nu}^{P (V_2)} &=& {^{*}\Gamma_{\mu \nu}}(K){\cal B}^0 +
   {{^{*}\Gamma}_{\mu}}^{\lambda}(K){\cal B}^1_{\lambda \nu}+
 {{^{*}\Gamma}_{\mu \lambda}}(K) \sum_{i}{\cal B}^{2}_{i\nu} P_i^{\lambda} 
       \nonumber \\
      & & + (\mu \leftrightarrow  \nu)~,
\label{PIV2}
\eea
where $(\mu \leftrightarrow  \nu)$ terms are the contributions from 
mirror diagrams.

The pinch part of the one-loop box diagrams depicted 
in Fig.6(b) has a form 
\be
  \widehat T_P^{(Box)}=[T^a] \biggl\{ [\gamma_{\alpha}] 
           [\gamma^{\alpha}]{\cal C}^0 + 
     \sum_{i,j} {\cal C}_{ij} [\Psl_i][\Psl_j]  \biggr\} [T^a].
\ee
Again from Eq.(\ref{Identity}) we see that 
$[\gamma_{\alpha}] [\gamma^{\alpha}]$ and $[\Psl_i][\Psl_j]$ are rewritten 
as     
\be
  [\gamma_{\alpha}] [\gamma^{\alpha}] =
     [\gamma_{\alpha}]~{^{*}D}^{\alpha \mu}(K)
           [~{^{*}\Gamma_{\mu \lambda}}(K)~{^{*}{\Gamma^{\lambda}}_{\nu}}(K)] 
           ~{^{*}D}^{\nu \beta}(K)[\gamma_{\beta}]              
\ee
\be
    [\Psl_i][\Psl_j] =
        [\gamma_{\alpha}]~{^{*}D}^{\alpha \mu}(K)[~{^{*}\Gamma_{\mu \lambda}}(K)
                ~{^{*}\Gamma_{\nu \tau}}(K) P_i^{\lambda} P_j^{\tau}]
     ~{^{*}D}^{\nu \beta}(K)[\gamma_{\beta}]
\ee
and thus we obtain for the contributions to  ${^{*}\Pi}^{\mu \nu}$ from the 
box diagrams
\be
   {^{*}\Pi}_{\mu \nu}^{P(Box)}=
   {^{*}\Gamma_{\mu \lambda}}(K)~{^{*}{\Gamma^{\lambda}}_{\nu}}(K){\cal C}^0 +
        {^{*}\Gamma_{\mu \lambda}}(K)
                ~{^{*}\Gamma_{\nu \tau}}(K) 
        \sum_{i,j}{\cal C}_{ij}P_i^{\lambda} P_j^{\tau}~.
\ee


\subsection{\it Resummed Pinch Contributions}
\smallskip

Following the prescription developed in Sec.4.1, we now obtain the resummed pinch 
contributions to the resummed gluon self-energy. First we present the results. 
The indivisual contributions from the vertex diagrams of the 
first kind, of the second kind and the box diagrams are presented 
in Appendix A.  In total they are expressed as 
\be
    {^{*}\Pi^{(P)}_{\mu\nu}}(K)=Ng^2 \int dP \bigl\{ 
        J^{(P)B}_{\mu\nu} + 
            \sum_{i\ge j} I^{(P)ij}_{\mu\nu}  \bigr\} 
    \qquad \qquad {\rm with} \quad i, j=A, B, S, T   
\ee 
where
\bea
 & & J^{(P)B}_{\mu\nu}=-B(P)\frac{1}{P^2}~{^{*}\Gamma_{\mu\nu}}(K)~,  
\label{IPB} \\
{\rm and} & & \nonumber  \\
 & & I^{(P)AA}_{\mu\nu}=  - A(P)A(Q) \Biggl[~2~{^{*}\Gamma_{\mu\nu}}(K)~ 
+ \biggl\{ {^{*}\Gamma_{\mu\alpha}}(K)~ {V^{\alpha}}_{\nu}(Q,K) 
  + (\mu \leftrightarrow \nu) \biggr\} \Biggr]~,  
\label{IPAA}  \\     
 & & I^{(P)BB}_{\mu\nu}=-\frac{1}{2} B(P)B(Q)\frac{1}{P^2 Q^2}
           P^{\lambda}P^{\tau}~{^{*}\Gamma_{\mu\lambda}}(K)~
                 {^{*}\Gamma_{\nu\tau}}(K)~,  
\label{IPBB}    \\
 & & I^{(P)SS}_{\mu\nu}=S(P)S(Q) \Biggl[~ -\frac{p_0 q_0}{P^2 Q^2} 
       K^4_L \widetilde n_{\mu}(K) \widetilde n_{\nu}(K) \nonumber  \\
  & &\qquad \qquad \qquad \qquad \quad + \frac{p_0}{P^2}
       \biggl\{ K^2_L \widetilde n_{\mu}(K) \widetilde n^{\alpha}(Q)~ 
          {^{*}\Gamma_{\alpha\nu 0}}(Q,K,P)
              + (\mu \leftrightarrow \nu)\biggr\} \nonumber \\
& &\qquad \qquad \qquad \qquad \quad - \frac{1}{2} 
    \frac{p_0 q_0}{P^2 Q^2} \biggl\{ P^{\lambda}~ 
   {^{*}\Gamma_{\lambda\mu}}(K)~{^{*}\Gamma_{0 \nu 0}}(Q,K,P)
 + (\mu \leftrightarrow \nu)\biggr\}
              \Biggr]~, 
\label{IPSS} \\
 & & I^{(P)TT}_{\mu\nu}=T(P)T(Q) 
       \biggl\{ K^4_L \widetilde n_{\mu}(K) \widetilde n_{\nu}(K) 
          - K^2_L P^2_L \widetilde n_{\mu}(P) \widetilde n_{\nu}(K) \nonumber  \\
     & & \qquad \qquad  - K^2_L Q^2_L \widetilde n_{\mu}(K) \widetilde n_{\nu}(Q) 
    -\frac{1}{2}\Bigl[
        P^{\lambda}~{^{*}\Gamma_{\lambda\mu}}(K)~ 
         {^{*}\Gamma_{0 \nu 0}}(Q,K,P) + (\mu \leftrightarrow \nu) \Bigr]
     \biggr\}~,  \\
 & & I^{(P)AB}_{\mu\nu}=A(Q)B(P)\frac{1}{P^2}
         \biggl\{ -{^{*}\Gamma_{\mu\alpha}}(K)~{^{*}\Gamma^{\alpha}}_{\nu}(K)
   + {^{*}\Gamma_{\mu\alpha}}(K)~{^{*}\Gamma^{\alpha}}_{\nu}(Q) 
          \nonumber \\
 & & \qquad \qquad \qquad \qquad \qquad \qquad \qquad \qquad \qquad \qquad 
   + {^{*}\Gamma_{\mu\alpha}}(Q)~{^{*}\Gamma^{\alpha}}_{\nu}(K)  \biggr\}~,  \\
 & & I^{(P)AS}_{\mu\nu}=A(Q)S(P) \Biggl[ \biggl\{ K^2_L \widetilde n_{\mu}(K)  
 \widetilde n^{\alpha}(P) \bigl[ g_{\alpha\nu}+V_{\alpha\nu}(Q,K) \bigr] 
+ (\mu \leftrightarrow \nu) \biggr\} \nonumber  \\
 & &\qquad \qquad \qquad \qquad \qquad - \frac{p_0}{P^2} \biggl\{ 
  ~{^{*}\Gamma_{\mu}}^{\alpha}(K)~{^{*}\Gamma_{\alpha\nu 0}}(Q,K,P)  
   + (\mu \leftrightarrow \nu) \biggr\}  
 \Biggr]~,  
\label{IPAS}\\
 & & I^{(P)AT}_{\mu\nu}= A(Q)T(P) \Biggl[ \biggl\{{^{*}\Gamma_{\mu\alpha}}(K)~ 
       {^{*}\Gamma^{\alpha}}_{\nu 0}(Q,K,P) + (\mu \leftrightarrow \nu) 
            \biggr\} \nonumber \\
 & &\qquad  + 
    \biggl\{ \Bigl[ (-Q^2_T+ Q^2_L) 
     \frac{q_0}{Q^2\widetilde n^2(Q)} \widetilde n_{\mu}(Q)
      +  Q^2_T \frac{ Q_{\mu}}{Q^2}  \Bigr]
        K^2_L  \widetilde n_{\nu}(K)  + (\mu \leftrightarrow \nu) 
         \biggr\}   \Biggr] , \\  
 & & I^{(P)BS}_{\mu\nu}=B(P)S(Q)\frac{1}{P^2}
   \Biggl[ -K^4_L \widetilde n_{\mu}(K) \widetilde n_{\nu}(K) 
     + \biggl\{ K^2_L\widetilde n_{\mu}(K) Q^2_L\widetilde n_{\nu}(Q) 
   + (\mu \leftrightarrow \nu) \biggr\}  \nonumber  \\
 & &\qquad \qquad \qquad - \frac{q_0}{Q^2} \biggl\{
   Q^{\lambda}~{^{*}\Gamma_{\lambda\mu}}(K)\Bigl[ 
   K^2_L\widetilde n_{\nu}(K) -  Q^2_L\widetilde n_{\nu}(Q) \Bigr]
      + (\mu \leftrightarrow \nu) \biggr\}   
       \Biggr]    \\
 & & I^{(P)BT}_{\mu\nu}=B(P)T(Q)\frac{1}{P^2}
         \biggl\{  
   \bigl[K^2_L \widetilde n_{\mu}(K) - Q^2_L \widetilde n_{\mu}(Q) \bigr] 
     Q^{\lambda}~{^{*}\Gamma_{\lambda\nu}}(K)     
      + (\mu \leftrightarrow \nu) \biggr\}    \\  
 & & I^{(P)ST}_{\mu\nu}=-S(Q)T(P) \Biggl[ 
         \biggl\{ \Bigl[ K^2_L \widetilde n_{\mu}(K) + 
       \frac{q_0}{Q^2} Q^{\lambda}~{^{*}\Gamma_{\lambda\mu}}(K) \Bigr]~
        {^{*}\Gamma_{0 \nu 0}}(Q,K,P) + (\mu \leftrightarrow \nu) \biggr\} 
          \nonumber  \\
 & & \quad  + \frac{q_0}{Q^2}\biggl[ 2 K^4_L \widetilde n_{\mu}(K) \widetilde n_{\nu}(K) 
     - \biggl\{K^2_L \widetilde n_{\mu}(K) \Bigl[ P^2_L \widetilde n_{\nu}(P) +
        Q^2_L \widetilde n_{\nu}(Q)
    \Bigr] + (\mu \leftrightarrow \nu) \biggr\}
          \biggr]  \Biggr] 
\label{IPST}
\eea
The function $V_{\alpha\nu}(P,K)$, which appeared in Eqs.(\ref{IPAA}) 
and (\ref{IPAS}), is defined  in Appendix B.

Now we explain how the above terms are obtained. 
Let us consider the pinch contribution from 
the vertex diagram of the second kind. The diagram of Fig.5(a) gives 
\bea 
\widehat T^{(V_2)} &=& \frac{N}{2}g^2\int dP~ [T^a]\biggl[\gamma _{\kappa} 
   \frac{1}{{\Lsl}-{\Psl}-m}\gamma_{\lambda} \biggr]
     ~{^{*}D}^{\lambda \tau}(P)~{^{*}D}^{\kappa \alpha}(Q)  \nonumber  \\
    & &  \qquad  \qquad  \qquad  \qquad  \qquad   
\times  ~{^{*}\Gamma_{\tau \nu \alpha}}(P,K,Q) 
        ~{^{*}D}^{\nu \beta}(K)[T^a \gamma_{\beta}]
\label{TPV2b}
\eea
The contribution of $\widehat T^{(V_2)}$ to $I^{(P)BB}_{\mu\nu}$ is 
extracted as follows. The product of two propagator 
${^{*}D}^{\lambda \tau}(P)~{^{*}D}^{\kappa \alpha}(Q)$ contains a term 
\be
          B(P)B(Q)\frac{P^{\lambda}P^{\tau}Q^{\kappa}Q^{\alpha}}{P^2 Q^2}~. 
\ee
According to the pinch prescription, the product 
\be
\biggl[\gamma _{\kappa} \frac{1}{{\Lsl}-{\Psl}-m}\gamma_{\lambda} \biggr] 
 P^{\lambda}Q^{\kappa}=
   [({\Lsl}-{\Psl}-m)-({\Lsl}+{\Ksl}-m)]\frac{1}{{\Lsl}-{\Psl}-m}{\Psl}
\ee
gives ${\Psl}$, since a term $({\Lsl}+{\Ksl}-m)$ vanishes on mass shell. 
Also due to the Ward-Takahashi identities (\ref{WTTwo}) and (\ref{WTThree}) 
satisfied by the effective two-point function and three-point vertex, we find 
\be
   P^{\tau}Q^{\alpha}~{^{*}\Gamma_{\tau \nu \alpha}}(P,K,Q) 
         = - P^{\tau}~{^{*}\Gamma_{\tau \nu }}(K)~.
\ee
Thus a component of $\widehat T^{(V_2)}_P$ which is relevant 
to $I^{(P)BB}_{\mu\nu}$ is written as
\bea
   \widehat T^{(V_2)}_P \Big{\vert}_{BB}&=&-\frac{N}{2}g^2\int 
           dP~ [T^a]B(P)B(Q)
      \frac{1}{P^2 Q^2}[{\Psl}]P^{\tau}~{^{*}\Gamma_{\tau \nu }}(K)  
      \nonumber  \\
& &  \qquad  \qquad  \qquad  \qquad \times {^{*}D}^{\nu \beta}(K) 
               [T^a \gamma_{\beta}]
\eea
The final step is to use Eqs.(\ref{TPV2}) and (\ref{PIV2}) 
and we obtain for the contribution of the vertex diagram of the second kind 
to $I^{(P)BB}_{\mu\nu}$, 
\be
   I^{(P)BB}_{\mu\nu}\Big{\vert}_{V_2}=-B(P)B(Q) \frac{1}{P^2 Q^2}
     ~{^{*}\Gamma_{\mu \lambda}}(K)P^{\lambda}P^{\tau}~{^{*}\Gamma_{\tau \nu }}(K)~, 
\ee
where we have added the mirror diagram contribution. 
The analysis of the pinch contributions from the box diagram can be done 
in a similar way and we find  $I^{(P)BB}_{\mu\nu}\Big{\vert}_{Box}=
 -\frac{1}{2}I^{(P)BB}_{\mu\nu}\Big{\vert}_{V_2}$. 
Then the sum of $I^{(P)BB}_{\mu\nu}\Big{\vert}_{V_2}$ and 
$I^{(P)BB}_{\mu\nu}\Big{\vert}_{Box}$ gives the expression in Eq.(\ref{IPBB}) for 
$I^{(P)BB}_{\mu\nu}$. 
The other terms in
Eq.(\ref{IPB}) through Eq.(\ref{IPST}) 
are obtained similarly, except for $I^{(P)AA}_{\mu\nu}$, 
$I^{(P)SS}_{\mu\nu}$ and $I^{(P)AS}_{\mu\nu}$. 

The derivation of $I^{(P)AA}_{\mu\nu}$, 
$I^{(P)SS}_{\mu\nu}$ and $I^{(P)AS}_{\mu\nu}$ are more involved.  
Let us start with $I^{(P)AA}_{\mu\nu}$ term. 
This term emerges from 
the vertex diagram of the second kind.
The product ${{^{*}D}_{\lambda}}^{\tau}(P)~{{^{*}D}_{\kappa}}^{\alpha}(Q)
~{^{*}\Gamma_{\tau \nu \alpha}}(P,K,Q)$ contains a term 
\be
       A(P)A(Q)~{^{*}\Gamma_{\lambda \nu \kappa}}(P,K,Q)~.
\ee
We expand ${^{*}\Gamma_{\lambda \nu \kappa}}(P,K,Q)$ into the sum of 
terms with different tensor structures, such as terms proportional 
to $P_{\lambda}$, terms proportional 
to $Q_{\kappa}$ and others. Then ${^{*}\Gamma_{\lambda \nu \kappa}}(P,K,Q)$ is 
rewritten as 
\be
  {^{*}\Gamma_{\lambda \nu \kappa}}(P,K,Q)=P_{\lambda} \bigl\{
  g_{\kappa \nu}+V_{\kappa \nu}(P,K) \bigr\} 
   - Q_{\kappa}\bigl\{ g_{\lambda \nu}+V_{\lambda \nu}(Q,K) \bigr\} 
  + \cdots 
\label{ExpGamA} 
\ee
where  the dots $\cdots$ represent terms which are neither 
proportional to $P_{\lambda}$ nor to $Q_{\kappa}$. The functions 
$V_{\kappa \nu}(P,K)$ and $V_{\lambda \nu}(Q,K)$ 
are given by Eqs.(\ref{Wexpand})-(\ref{Vexpand}) 
in Appendix B. It is symmetric in indices $\lambda$ and $\nu$ and 
satisfies the following identity (see also Eq.(\ref{IdentityV})):
\be
          K^{\nu}V_{\lambda \nu}(Q,K)=\Bigl( 1-\frac{Q_T^2}{Q^2} \Bigr) 
           Q_{\lambda} + \Bigl( Q_T^2 -Q_L^2 \Bigr) 
        \frac{q_0}{Q^2} \frac{\widetilde n_{\lambda}(Q)}
          {\widetilde n^2(Q)}~.
\label{WardV}
\ee
Now the pinch prescription gives 
\be
   \gamma^{\kappa}\frac{1}{{\Lsl}-{\Psl}-m} {\Psl} \Longrightarrow - \gamma^{\kappa}~,
\ee
\be
  -{\Qsl}\frac{1}{{\Lsl}-{\Psl}-m} \gamma^{\lambda} \Longrightarrow - \gamma^{\lambda}~.
\ee
Thus a pinch part $\widehat T^{(V_2)}_P$ which is relevant 
to $I^{(P)AA}_{\mu\nu}$ is expressed as 
\bea
   \widehat T^{(V_2)}_P \Big{\vert}_{AA}&=&-N g^2\int dP~ [T^a]A(P)A(Q)~
           \bigl[\gamma^{\lambda}\bigr] 
        \bigl[g_{\lambda \nu} +V_{\lambda \nu}(Q,K) \bigr] 
      \nonumber  \\
& &  \qquad  \qquad  \qquad  \qquad \qquad \times {^{*}D}^{\nu \beta}(K)~ 
               [T^a \gamma_{\beta}]~,
\label{TV2PAA}
\eea 
where we have used the symmetry property of the integrand under the interchange of 
$P$ and $Q$.  The mirror diagram contribution is obtained by interchanging 
indices $\mu$ and $\nu$ in the above expression of 
$\widehat T^{(V_2)}_P \Big{\vert}_{AA}$.
Then the formulas (\ref{TPV2}) and (\ref{PIV2})   
give $I^{(P)AA}_{\mu\nu}$ in Eq.(\ref{IPAA}).

Now we proceed to the derivation of $I^{(P)AS}_{\mu\nu}$ term, 
which emerges from the vertex diagram of the second kind. It is reminded 
that the $S$ term in the resummed gluon propagator is typical at finite 
temperature. Indeed the $S$ term will vanish  at $T=0$.  
The product ${{^{*}D}_{\lambda}}^{\tau}(P)~{{^{*}D}_{\kappa}}^{\alpha}(Q)
 ~{^{*}\Gamma_{\tau \nu \alpha}}(P,K,Q)$  in Eq.(\ref{TPV2b}) contains terms 
\bea
  & & A(Q)S(P)\Bigl\{n_{\lambda}n^{\tau}-\frac{p_0}{P^2}
   \bigl[n_{\lambda}P^{\tau}+P_{\lambda}n^{\tau}\bigr] \Bigr\}
    ~{^{*}\Gamma_{\tau \nu \kappa}}(P,K,Q)   \nonumber  \\
  &+&A(P)S(Q)\Bigl\{n_{\kappa}n^{\alpha}-\frac{q_0}{Q^2}
   \bigl[n_{\kappa}Q^{\alpha}+Q_{\kappa}n^{\alpha}\bigr] \Bigr\}
    ~{^{*}\Gamma_{\lambda \nu \alpha}}(P,K,Q) ~. 
\label{ASterm}
\eea
The second line gives the same contribution as the first one. 
The first line is rewritten as 
\bea
& & A(Q)S(P)\Bigl\{n_{\lambda}
\widetilde n^{\tau}(P)~{^{*}\Gamma_{\tau \nu \kappa}}(P,K,Q)
  -\frac{p_0}{P^2}P_{\lambda} ~{^{*}\Gamma_{0 \nu \kappa}}(P,K,Q)\Bigr\} 
\nonumber  \\
& & \qquad = A(Q)S(P)\biggl[n_{\lambda}
\widetilde n^{\tau}(P) \Bigl\{ -Q_{\kappa}
 \bigl[g_{\tau\nu} + V_{\tau \nu}(Q,K) \bigr]  \Bigr\} \nonumber  \\
& &  \qquad \qquad \qquad \qquad \qquad 
  -\frac{p_0}{P^2}P_{\lambda} ~{^{*}\Gamma_{0 \nu \kappa}}(P,K,Q)
     + \cdots \biggr]~,
\eea
where a decomposed form for ${^{*}\Gamma_{\tau \nu \kappa}}(P,K,Q)$ 
such as given in Eq.(\ref{ExpGamA}) and an identity 
$\widetilde n^{\tau}(P) P_{\tau}=0$ were used. 
The dots $\cdots$ represent irrelevant terms which are not 
proportional to $P_{\lambda}$ nor to $Q_{\kappa}$ and, therefore, 
do not yield the pinch parts. Then the pinch technique prescription gives 
a pinch part $\widehat T^{(V_2)}_P \Big{\vert}_{AS}$ 
which is relevant to $I^{(P)AS}_{\mu\nu}$ as follows: 
\bea
   \widehat T^{(V_2)}_P \Big{\vert}_{AS}&=&N g^2\int dP~ [T^a]A(Q)S(P)~
           \biggl\{-\bigl[\nsl \bigr] \widetilde n^{\tau}(P) 
    \bigl[g_{\tau\nu} + V_{\tau \nu}(Q,K) \bigr]  \nonumber  \\
& & \qquad  \qquad  \qquad  \qquad + 
      \bigl[\gamma^{\kappa}\bigr] \frac{p_0}{P^2}
         ~{^{*}\Gamma_{0 \nu \kappa}}(P,K,Q) \biggr\}~,
\label{TV2PAS}
\eea 
where the contribution of the second line in Eq.(\ref{ASterm}) has been  added.
Now it is straightforward to obtain the expression of 
Eq.(\ref{IPAS}) for $I^{(P)AS}_{\mu\nu}$. 
The derivation of  
$I^{(P)SS}_{\mu\nu}$ can be done in a similar way.

\bigskip
\section{Gauge-independent Resummed Gluon Self-Energy}
\smallskip

In this section we show that when we combine the resummed gluon 
self-energy calculated in Sect.3 with the resummed pinch contributions 
in Sect.4, we will obtain the effective gluon self-energy 
which is {\it gauge independent} and also {\it satisfies the transversality 
relation}.

\bigskip
\subsection{\it Gauge-independence}
\smallskip
First we show that once the pinch contributions are added, the $B$-terms 
are cancelled out. From the expressions for the $B$-terms obtained in 
Sect.3 and Sect.4, we see that
\bea
          I^{BB}_{\mu\nu}+I^{(P)BB}_{\mu\nu} &=& 0  \\
          I^{BT}_{\mu\nu}+I^{(P)BT}_{\mu\nu} &=& 0  
\eea
and
\be
     I^{AB}_{\mu\nu}+I^{(P)AB}_{\mu\nu}+I^{BS}_{\mu\nu}+I^{(P)BS}_{\mu\nu}
   = B(P)\frac{1}{P^2}{^{*}\Gamma_{\mu\nu}}(Q) 
\ee
On the other hand, we find  
\be
    J^{B}_{\mu\nu}+J^{(P)B}_{\mu\nu}=-B(P)\frac{1}{P^2}{^{*}\Gamma_{\mu\nu}}(Q)
\ee
Thus, all the $B$-terms, when added together, cancel out, which means 
that the gauge patrameter $\xi$-dependence disappears.

Next we show that the $T$-terms, when the pinch contributions are added,  
also cancel out and thus all the gauge-dependent terms disappear. 
First we have  
\bea
   I^{TT}_{\mu\nu}+I^{(P)TT}_{\mu\nu}&=& -T(P)T(Q)
     P^2_L Q^2_L \widetilde n_{\mu}(P)\widetilde n_{\nu}(Q) \nonumber \\
     &=& -\frac{1}{P^2Q^2}\frac{b(P)b(Q)}{a(P)a(Q)}
           \widetilde n_{\mu}(P)\widetilde n_{\nu}(Q)
\label{SumTT}
\eea
where, in the second line, Eq.(\ref{TTT}) has been used for $T(P)T(Q)$. 
It is noted that the sum of $I^{TT}_{\mu\nu}+I^{(P)TT}_{\mu\nu}$  cancels 
against a term proportinal to $\frac{b(P)b(Q)}{a(P)a(Q)}$ 
in ${\Pi^{gh}_{\mu\nu}}$ given in Eq.(\ref{PiGhost}). 
Using the decomposed expression  
\bea
  {^{*}\Gamma_{\mu\alpha}}(Q)~{^{*}\Gamma^{\alpha}}_{\nu 0}(Q,K,P) &=& 
      - Q^2_T \biggl\{{^{*}\Gamma_{\mu\nu 0}}(Q,K,P) +
         \frac{Q_{\mu}}{Q^2} \Bigl[ K^2_L \widetilde n_{\nu}(K) -  
     P^2_L \widetilde n_{\nu}(P) \Bigr] \biggr\} \nonumber \\
  & & +(Q^2_T-Q^2_L)\frac{\widetilde n_{\mu}(Q)}{\widetilde n^2(Q)}
   \biggl\{{^{*}\Gamma_{0 \nu 0}}(Q,K,P)   \nonumber \\
  & & \qquad \qquad \qquad \qquad \ 
       +  \frac{q_0}{Q^2} \Bigl[ K^2_L \widetilde n_{\nu}(K) - 
     P^2_L \widetilde n_{\nu}(P) \Bigr] \biggr\} 
\eea
we find that the sum of $I^{AT}_{\mu\nu}+I^{(P)AT}_{\mu\nu}$ is written as 
\bea
  I^{AT}_{\mu\nu}+I^{(P)AT}_{\mu\nu}&=& 
        T(P)\Biggl[ \biggl\{ {^{*}\Gamma_{\mu\nu 0}}(Q,K,P) + 
              (\mu \leftrightarrow \nu) \biggr\} \nonumber \\
 & & \qquad \qquad -  \Bigl(1- \frac{Q^2_L}{Q^2_T} \Bigr) 
       \frac{1}{\widetilde n^2(Q)} \Bigl\{ \widetilde n_{\mu}(Q) 
  {^{*}\Gamma_{0 \nu 0}}(Q,K,P) + (\mu \leftrightarrow \nu) \Bigr\}
                \nonumber \\
 & & \qquad \qquad  + \Bigl(1- \frac{Q^2_L}{Q^2_T} \Bigr) 
      \frac{q_0}{Q^2 \widetilde n^2(Q)}  P^2_L 
  \Bigl\{ \widetilde n_{\mu}(Q) \widetilde n_{\nu}(P) + 
     (\mu \leftrightarrow \nu)  \Bigr\}  \nonumber \\
 & & \qquad \qquad 
       - \frac{P^2_L}{Q^2} \Bigl\{Q_{\mu} \widetilde n_{\nu}(P) 
       + (\mu \leftrightarrow \nu) \Bigr\} 
  \Biggr] 
\label{SumAT}
\eea
where we have used Eq.(\ref{SSS}) for $S(Q)$.
The first line cancels against $J^T_{\mu\nu}$ and the second and third lines  
cancel against the sum of $I^{ST}_{\mu\nu}$ and $I^{(P)ST}_{\mu\nu}$. 
Thus we have 
\bea
 & & \bigl[I^{AT}_{\mu\nu}+I^{(P)AT}_{\mu\nu}\bigr]+ 
       \bigl[I^{ST}_{\mu\nu}+I^{(P)ST}_{\mu\nu}\bigr] 
+ J^T_{\mu\nu} \nonumber \\
 & & \qquad \qquad = -T(P)  \frac{P^2_L }{Q^2}\Bigl\{ 
       Q_{\mu} \widetilde n_{\nu}(P) 
              + (\mu \leftrightarrow \nu) \Bigr\} \nonumber  \\
 & & \qquad \qquad = - \frac{1}{P^2 Q^2}\frac{b(P)}{a(P)} 
   \Bigl\{  Q_{\mu} \widetilde n_{\nu}(P) + \widetilde n_{\mu}(P) Q_{\nu} 
  \Bigr\}~,
\eea
which cancels against a term proportinal to $\frac{b(P)}{a(P)}$
in ${\Pi^{gh}_{\mu\nu}}(K)$ given by  Eq.(\ref{PiGhost}).  
Therefore, the $a$- and $b$-dependence of the $T$-terms and  
${\Pi^{gh}_{\mu\nu}}(K)$ completely cancel out.

Summing up the remaining terms, we find for the gauge-independent 
resummed gluon self-energy, 
\bea
    {^{*}\widehat{\Pi}_{\mu\nu}}(K)&=&
     {^{*}\Pi_{\mu\nu}}(K) + {^{*}\Pi^P_{\mu\nu}}(K) \nonumber \\
    &=& Ng^2 \int dP \biggl\{ 
     \bigl[I^{AA}_{\mu\nu}+I^{SS}_{\mu\nu}+I^{AS}_{\mu\nu}\bigr] + 
     \bigl[J^{A}_{\mu\nu}+J^{S}_{\mu\nu}\bigr] \nonumber \\ 
 & & \qquad  \qquad \quad +\bigl[I^{(P)AA}_{\mu\nu}+I^{(P)SS}_{\mu\nu}+
     I^{(P)AS}_{\mu\nu} \bigr] 
         + I^{rest}_{\mu\nu} \biggr\} 
\eea
with
\be
  I^{rest}_{\mu\nu}=\frac{1}{P^2 Q^2}P_{\mu}Q_{\nu}.
\ee
where $I^{AA}_{\mu\nu}$, $I^{SS}_{\mu\nu}$, $I^{AS}_{\mu\nu}$, $J^{A}_{\mu\nu}$, 
$I^{S}_{\mu\nu}$, $I^{(P)AA}_{\mu\nu}$, $I^{(P)SS}_{\mu\nu}$, and 
$I^{(P)AS}_{\mu\nu}$ are given in Eqs.(\ref{IAA}), (\ref{ISS}), (\ref{IAS}), 
(\ref{JA}), (\ref{JS}), (\ref{IPAA}), (\ref{IPSS}), and (\ref{IPAS}), 
respectively. Explicitly, ${^{*}\widehat{\Pi}_{\mu\nu}}(K)$ is written as 
\bea
   {^{*}\widehat{\Pi}_{\mu\nu}}(K)&=&Ng^2 \int dP \Biggl\{~ 
       - A(P)A(Q)~\biggl[~\frac{1}{2}~
            {^{*}\Gamma_{\lambda\mu}}^{\alpha}(P,K,Q)~ 
            {^{*}\Gamma_{\alpha\nu}}^{\lambda}(Q,K,P) \nonumber  \\
 & & \qquad \qquad \qquad \qquad + ~2~{^{*}\Gamma_{\mu\nu}}(K)~
  + \Bigl\{ {^{*}\Gamma_{\mu\alpha}}(K)~ {V^{\alpha}}_{\nu}(Q,K) 
  + (\mu \leftrightarrow \nu) \Bigr\} \biggr] \nonumber \\
 & & -S(P)S(Q)~\biggl[~ \frac{1}{2}~
           {^{*}\Gamma_{0 \mu 0}}(P,K,Q)~{^{*}\Gamma_{0 \nu 0}}(Q,K,P) 
   \nonumber   \\
  & & \qquad  \qquad \qquad +\frac{p_0q_0}{P^2Q^2} Q^2_L 
   \Bigl\{~\widetilde n_{\mu}(Q)~K^2_L \widetilde n_{\nu}(K) 
    +   (\mu \leftrightarrow \nu) \Bigr\} \nonumber  \\
   & & \qquad  \qquad \qquad -\frac{p_0q_0}{P^2Q^2} P^2_L 
   \widetilde n_{\mu}(P)~Q^2_L \widetilde n_{\nu}(Q) 
   \biggr] \nonumber  \\
& & -A(Q)S(P)~\biggl[~{^{*}\Gamma_{0 \mu \alpha }}(P,K,Q)~
              {^{*}\Gamma^{\alpha}}_{\nu 0}(Q,K,P) \nonumber  \\
& & \qquad  \qquad \qquad - \Bigl\{ K^2_L \widetilde n_{\mu}(K)  
 \widetilde n^{\alpha}(P)~ \bigl[ g_{\alpha\nu}+V_{\alpha\nu}(Q,K) \bigr] 
+ (\mu \leftrightarrow \nu) \Bigr\} \biggr] \nonumber  \\
 & & + \frac{1}{2}A(P)~
        {^{*}\Gamma_{\mu\nu\alpha}}^{\alpha}(K,-K,P,-P) 
 +  \frac{1}{2}S(P)~{^{*}\Gamma_{\mu\nu 0 0}}(K,-K,P,-P) \nonumber  \\
& & -S(P)~\frac{p_0}{P^2Q^2}\Bigl\{ Q_{\mu}
   \Bigl[ K^2_L \widetilde n_{\nu}(K) -  P^2_L \widetilde n_{\nu}(P) \Bigr]
   + (\mu \leftrightarrow \nu) \Bigr\}  \nonumber \\
& & + \frac{1}{P^2 Q^2}P_{\mu}Q_{\nu} \Biggr\}~.
\label{InvPi}
\eea
Since ${^{*}\widehat{\Pi}_{\mu\nu}}(K)$ does not depend on $a$, $b$ and $\xi$, 
it is clear that ${^{*}\widehat{\Pi}_{\mu\nu}}(K)$ is gauge independent. 
Also it is emphasized that inclusion of pinch contributions,  
$I^{(P)AA}_{\mu\nu}$, $I^{(P)SS}_{\mu\nu}$, and $I^{(P)AS}_{\mu\nu}$, 
to ${^{*}\widehat{\Pi}_{\mu\nu}}(K)$ is indispensable. Otherwise, 
${^{*}\widehat{\Pi}_{\mu\nu}}(K)$ does not satisfy the transversality relation, 
which will be shown in the next subsection.

\bigskip
\subsection{Transversality of ${^{*}\widehat{\Pi}_{\mu\nu}}(K)$}
\smallskip

The effective resummed gluon self-energy ${^{*}\widehat{\Pi}_{\mu\nu}}(K)$ 
obtained in the above subsection is not only gauge-independent, but also 
it satisfies the transversality relation
\be
  K^{\mu}~{^{*}\widehat{\Pi}_{\mu\nu}}(K) = 0
\label{Transversality}
\ee
This can be shown by explicit calculation. 
By applying $K^{\mu}$ to the sum $(I^{AA}_{\mu\nu}+I^{SS}_{\mu\nu}+
I^{AS}_{\mu\nu}+J^{A}_{\mu\nu}+J^{S}_{\mu\nu}+I^{rest}_{\mu\nu})$, we find 
\bea
& &K^{\mu}\bigl[(I^{AA}_{\mu\nu}+I^{SS}_{\mu\nu}+I^{AS}_{\mu\nu})+ 
    (J^{A}_{\mu\nu}+J^{S}_{\mu\nu})+I^{rest}_{\mu\nu}\bigr]  \nonumber \\
& & \qquad = A(P)\frac{Q^{\lambda}}{Q^2}~{^{*}\Gamma_{\lambda \nu}}(K)
      \nonumber \\
& & \qquad \quad - S(P)\biggl\{ \frac{Q\cdot \widetilde n(P) }{Q^2} 
      K_L^2 \widetilde n_{\nu}(K) + \frac{p_0 q_0}{P^2 Q^2}
           P^{\lambda}~{^{*}\Gamma_{\lambda \nu}}(K) \biggr\}
\label{TransA}
\eea
where we have discarded terms which are odd functions in $P$.

On the other hand, using the identity Eq.(\ref{WardV}) satisfied by the function 
${V^{\alpha}}_{\mu}(Q,K)$, we find   
\bea
  K^{\mu}I^{(P)AA}_{\mu\nu}&=&A(P)
  \frac{Q^{\lambda}}{Q_T^2}~{^{*}\Gamma_{\lambda \nu}}(K) \nonumber \\
  &-& A(P)\frac{Q^{\lambda}}{Q^2}~{^{*}\Gamma_{\lambda \nu}}(K) \nonumber \\
  &+& A(P)\bigl(1-\frac{Q_L^2}{Q_T^2}\bigr)
      \frac{1}{\widetilde n^2(Q)}\frac{q_0}{Q^2}
        \widetilde n^{\lambda}(Q)~{^{*}\Gamma_{\lambda \nu}}(K). 
\label{TransAA}
\eea
The first term turns out to vanish
due to the symmetry property of the integrand under the interchange of 
$P$ and $Q$, i.e., 
\bea
  & & A(P) \frac{Q^{\lambda}}{Q_T^2}~{^{*}\Gamma_{\lambda \mu}}(K) 
       \Longrightarrow -\frac{1}{2}\frac{P^{\lambda}+Q^{\lambda}}{P_T^2 Q_T^2}
             ~{^{*}\Gamma_{\lambda \mu}}(K)  \nonumber  \\
 & & \qquad \qquad \qquad \qquad \qquad =
    \frac{1}{2}\frac{1}{P_T^2 Q_T^2}K^{\lambda}
             ~{^{*}\Gamma_{\mu \lambda}}(K) = 0~.
\eea

In a similar way we obtain  
\bea 
  K^{\mu}I^{(P)SS}_{\mu\nu}
   &=&S(P)S(Q) \biggl\{\frac{p_0}{P^2}\Bigl[ Q_L^2 \widetilde n^2(Q)- 
    P_L^2 \widetilde n(P)\cdot \widetilde n(Q)~ \Bigr] 
        K_L^2 \widetilde n_{\nu}(K) \nonumber  \\
   & & \qquad \qquad \qquad  + \frac{p_0 q_0}{P^2 Q^2}
       P_L^2 \widetilde n^2(P) P^{\lambda}~{^{*}\Gamma_{\lambda \nu}}(K) 
  \biggr\}~,    
\label{TransSS} 
\\
  K^{\mu}I^{(P)AS}_{\mu\nu}
   &=&A(Q)S(P) \biggl\{\Bigl[-\frac{Q_T^2}{Q^2}Q\cdot \widetilde n(P) 
   +(Q_T^2-Q_L^2)\frac{q_0}{Q^2} 
  \frac{\widetilde n(P)\cdot \widetilde n(Q)}{\widetilde n(Q)^2}~  
  \Bigr] K_L^2 \widetilde n_{\nu}(K) 
\nonumber  \\
& & \qquad \qquad \qquad  + \frac{p_0}{P^2}\Bigl[ P_L^2 \widetilde n^{\lambda}(P)- 
    Q_L^2 \widetilde n^{\lambda}(Q) \Bigr] 
  ~ {^{*}\Gamma_{\lambda \nu}}(K)  
  \biggr\}~.
\label{TransAS}
\eea
Then the sum of Eqs.(\ref{TransAA}), (\ref{TransSS}) and (\ref{TransAS}) becomes 
\be
 -A(P)\frac{Q^{\lambda}}{Q^2}~{^{*}\Gamma_{\lambda \nu}}(K)
  + S(P)\biggl\{ \frac{Q\cdot \widetilde n(P) }{Q^2} 
      K_L^2 \widetilde n_{\nu}(K) + \frac{p_0 q_0}{P^2 Q^2}
           P^{\lambda}~{^{*}\Gamma_{\lambda \nu}}(K) \biggr\}~, 
\ee
which is just an opposite of Eq.(\ref{TransA}). Thus we reach the result of 
Eq.(\ref{Transversality}).

Since the effective resummed gluon self-energy 
${^{*}\widehat{\Pi}_{\mu\nu}}(K)$ 
satisfies the transversality relation, it can be  
decomposed as 
\be
   {^{*}\widehat{\Pi}}_{\mu\nu}(K)=
      {^{*}\widehat{\Pi}_{\perp}}(K)P_{\mu\nu}(K)+
      {^{*}\widehat{\Pi}_{\parallel}}(K)Q_{\mu\nu}(K) 
\ee
where ${^{*}\widehat{\Pi}}_{\perp}(K)$ and 
${^{*}\widehat{\Pi}_{\parallel}}(K)$ 
are refered to the transverse and longitudinal functions, respectively.
Both functions can be extracted by applying the projection operators 
to ${^{*}\widehat{\Pi}}_{\mu\nu}(K)$ as follows:
\bea
  {^{*}\widehat{\Pi}}_{\perp}(K)&=&\frac{1}{2}P^{\mu\nu}(K)~ 
               {^{*}\widehat{\Pi}}_{\mu\nu}(K)   \\
  {^{*}\widehat{\Pi}_{\parallel}}(K)&=&
         Q^{\mu\nu}(K)~{^{*}\widehat{\Pi}}_{\mu\nu}(K)~. 
\eea

To show the usefulness of the effective resummed gluon self-energy 
$ {^{*}\widehat{\Pi}}_{\mu\nu}(K)$ obtained here, we 
will calculate in the next section the dampimg rate for the transverse gluons 
in the leading order. 


\bigskip
\section{Gluon Damping Rate in the Leading Order}
\smallskip


There had been much controversy over the gluon damping rates 
in hot QCD. ``Naive" one-loop calculations for the damping rate of gluons  
showed that it is gauge dependent in both magnitude and sign. 
It was then realized~\cite{Pisa} that there are higher-loop diagrams which 
contribute to the same order in the coupling constant $g$ 
as the one-loop diagram. These higher-order effects were  
resummed into effective propapagators and vertices in a systematic way. 
The complete calculation in the leading order 
was first made by Braaten and Pisarski~\cite{BPa}. 
As we already know, the resummed gluon self-energy 
$^{*}\Pi ^{\mu\nu}$ contains gauge-dependent and gauge-independent pieces. 
To isolate the gauge-independent pieces in $^{*}\Pi ^{\mu\nu}$ and to 
calculate the gluon damping rate, Braaten and Pisarski constructed the 
two-gluon $\cal T$-matrix element 
by putting $^{*}\Pi ^{\mu\nu}$ on the mass-shell and sandwiching it between 
physical wave functions.  The gluon damping rate came from the 
imaginary part of the $\cal T$ and the actual calculation was made in 
the Coulomb gauge.

We now calculate the damping rate for transverse gluons at zero momentum 
in a different way from the one taken by Braaten and Pisarski~\cite{BPa}.   
In the previous section, we have constructed the 
gauge-independent resummed gluon self-energy 
${^{*}\widehat{\Pi}}_{\mu\nu}(K)$ in the leading order. 
Since ${^{*}\widehat{\Pi}}_{\mu\nu}(K)$ satisfies the transversality 
relation, it is written in terms of the transverse function 
${^{*}\widehat{\Pi}_{\perp}(K)}$  and the longitudinal function
${^{*}\widehat{\Pi}_{\parallel}(K)}$. Then,  
in the leading order in $g$, the damping rate $\gamma_t (0)$ 
for the transverse gluon at zero momentum is given by 
\be
   \gamma_t (0) = \frac{1}{4 m_g} {\rm Disc} 
             {^{*}\widehat{\Pi}_{\perp} }(k_0=m_g, {\bf k}=0)~.
\ee
where $m_g=\frac{1}{3}\sqrt{N}gT$ is the gluon mass induced by the 
thermal medium.

In the limit ${\bf k} \rightarrow 0$, it follows 
from $O(3)$ rotational invariance that 
$^{*}\widehat{\Pi}_{ij}= {^{*}\widehat{\Pi}_{\perp}} \delta_{ij}$ so that 
\be
     {^{*}\widehat{\Pi}_{\perp} }(k_0, {\bf k}=0) = \frac{1}{3}
        {^{*}\widehat{\Pi}_{ii}} (k_0, {\bf k}=0)     
\ee
To evaluate ${^{*}\widehat{\Pi}_{ii}} (k_0, {\bf k}=0)$, we require 
the information on the effective vertices which appear in the sum 
\bea
   & &-\frac{1}{2} A(P)A(Q)~
            {^{*}\Gamma_{\lambda i}}^{\alpha}(P,K,Q)~ 
            {^{*}\Gamma_{\alpha i}}^{\lambda}(Q,K,P)  \nonumber  \\
   & &
   - \frac{1}{2}S(P)S(Q)~
           {^{*}\Gamma_{0 i 0}}(P,K,Q)~{^{*}\Gamma_{0 i 0}}(Q,K,P) 
    \nonumber  \\
  & & -A(Q)S(P)~{^{*}\Gamma_{0 i \alpha }}(P,K,Q)~
              {^{*}\Gamma^{\alpha}}_{i 0}(Q,K,P)
\nonumber \\
   & & \qquad \qquad = \frac{1}{2}\frac{1}{{\bf p}^2 {\bf q}^2} 
    \biggl[ \frac{p_0^2}{P_T^2}-\frac{P^2}{P_L^2} \biggr] 
    \biggl[ \frac{q_0^2}{Q_T^2}-\frac{Q^2}{Q_L^2} \biggr] 
     {^{*}\Gamma_{00i}}(P,Q,K)~{^{*}\Gamma_{00i}}(P,Q,K) \nonumber \\
  & & \qquad \qquad \ - \frac{1}{{\bf p}^2  Q_T^2} \biggl[ \frac{p_0^2}{P_T^2}    
            - \frac{P^2}{P_L^2}  \biggr] 
    {^{*}\Gamma_{0ji}}(P,Q,K)~{^{*}\Gamma_{0ji}}(P,Q,K) \nonumber \\ 
  & & \qquad \qquad \ + \frac{1}{2}  \frac{1}{P_T^2 Q_T^2} 
      {^{*}\Gamma_{lji}}(P,Q,K)~{^{*}\Gamma_{lji}}(P,Q,K)~,     
\eea
and 
\be
   J_{ii}^{S}= -\frac{1}{2}\biggl[\frac{1}{P_T^2}-\frac{1}{P_L^2} \biggr]
      \frac{P^2}{{\bf p}^2}~{^{*}\Gamma_{ii00}}(K,-K,P,-P)~. 
\ee
The products of the effective three-gluon vertices 
${^{*}\Gamma_{00i}}{^{*}\Gamma_{00i}}$, ${^{*}\Gamma_{0ji}}{^{*}\Gamma_{0ji}}$, 
and ${^{*}\Gamma_{lji}}{^{*}\Gamma_{lji}}$
and also the four-gluon vertex ${^{*}\Gamma_{ii00}}$, all of which are evaluated at 
${\bf k}=0$ and $k_0=m_g$, are given in Appendix C. 
Using these expressions, we find for the discontinuity 
coming from the sum of  $I^{AA}_{ii}$, $I^{SS}_{ii}$, and $I^{AS}_{ii}$ 
at ${\bf k}=0$ and $k_0=m_g$, 
\bea
 & &{\rm Disc} \int dP \bigl[ I^{AA}_{ii}+I^{AS}_{ii}+I^{SS}_{ii} \bigr] 
   \Big\vert_{{\bf k}=0, k_0=m_g} \nonumber \\
 & & \qquad  = {\rm Disc} \int dP \Biggl[ 
      \frac{1}{P_T^2 Q_T^2}\frac{9}{{\bf p}^2}({\bf p}^2 + p_0 q_0 )^2 
  + \frac{1}{P_L^2 Q_T^2}\frac{P^2}{{\bf p}^2} \frac{9 Q^4}{2 m_g^2} 
  + \frac{1}{P_L^2 Q_L^2}\frac{P^2 Q^2}{{\bf p}^2 {\bf q}^2} \frac{9 {\bf p}^2}{2}
          \nonumber  \\
   & & \qquad \qquad
  - \frac{P_L^2}{Q_T^2} \biggl[ \frac{Q^2}{P^2} + \frac{P^2}{2{\bf p}^2} 
                 \biggr] 
        \frac{1}{m_g^2} + \frac{Q_L^2}{P_L^2} \frac{P^2}{Q^2}\frac{1}{m_g^2} 
   + \frac{{\bf p}^2}{P^2 Q^2}
\Biggr]~.
\eea
The contributions of $J_{ii}^{A}$, $J_{ii}^{S}$A, and $I_{ii}^{rest}$ to 
the discontinuity at  ${\bf k}=0$ and $k_0=m_g$ are, respectively, 
\bea
  {\rm Disc} \int dP J_{ii}^{A} \Big\vert_{{\bf k}=0, k_0=m_g} &=& 0   \\
  {\rm Disc} \int dP J_{ii}^{S} \Big\vert_{{\bf k}=0, k_0=m_g} &=& 
    {\rm Disc} \int dP \Biggl[ \frac{P_L^2}{Q_T^2}\frac{Q^2}{P^2} \frac{1}{m_g^2} 
       - \frac{Q_L^2}{P_L^2}\frac{P^2}{Q^2} \frac{1}{m_g^2} \Biggr]~  \\
 {\rm Disc} \int dP I_{ii}^{rest} \Big\vert_{{\bf k}=0, k_0=m_g} &=& 
   {\rm Disc} \int dP \Biggl[ - \frac{{\bf p}^2}{P^2 Q^2}
      \Biggr]  
\eea
Finally the pinch contributions $I^{(P)AA}_{ii}$, $I^{(P)SS}_{ii}$, 
and $I^{(P)AS}_{ii}$ 
do not contribute, because they vanish at $k_0=m_g$ and ${\bf k}=0$.

Summing up each contribution, the discontinuity in the 
transverse function is written as 
\bea
  & &{\rm Disc}{^{*}\widehat{\Pi}_{\perp}}(k_0=m_g, {\bf k}=0) \nonumber \\
  & & \qquad  = 
  \frac{3}{2}g^2 N {\rm Disc} \int dP \Biggl[ 
     \frac{1}{P_T^2 Q_T^2}\frac{2}{{\bf p}^2} ({\bf p}^2 + p_0 q_0 )^2 
  + \frac{1}{P_L^2 Q_T^2}\frac{P^2}{{\bf p}^2} \frac{Q^4}{m_g^2} 
  \nonumber \\
  & & \qquad  \qquad \qquad \qquad \qquad
  + \frac{1}{P_L^2 Q_L^2}\frac{P^2 Q^2}{{\bf p}^2 {\bf q}^2} {\bf p}^2
  - \frac{P_L^2}{Q_T^2} \frac{P^2}{{\bf p}^2}  
        \frac{1}{9m_g^2} \Biggr]
\label{Disc}
\eea
Since $P_L^2=P^2-\delta \Pi_L (P)$ and 
\be
  \delta \Pi_L (P) = \frac{3m_g^2 P^2}{{\bf p}^2}
    \biggl[ \frac{1}{2}\frac{p_0}{p} {\rm ln}\frac{p_0 + p}{p_0 - p} - 1 \biggr]~,
\ee
the fourth term in Eq.(\ref{Disc}) is rewritten as 
\bea
 & &{\rm Disc} \int dP \biggl[- \frac{P_L^2}{Q_T^2} \frac{P^2}{{\bf p}^2}  
        \frac{1}{9m_g^2} \biggr]  \nonumber \\
 & & \qquad \qquad = {\rm Disc} \int dP \Biggl[\frac{1}{Q_T^2}
        \frac{1}{3} \frac{P^4 p_0}{p^5} \biggl\{ 
        \frac{1}{2} {\rm ln}\frac{p_0 + p}{p_0 - p} \biggr\}  \Biggr]
\eea
Thus we find that the expression of 
${\rm Disc}{^{*}\widehat{\Pi}_{\perp}}$ in Eq.(\ref{Disc}) is 
equivalent to the one\footnote{A factor $\frac{1}{3}$ is missing in the last term,  
which is proportional to $Q_0\Bigl(\frac{ik^0}{k}\Bigr)$,  
in Eq.(22) of Ref.\cite{BPb}. It is only a typographical error and does not
propagate to Eq.(25), and thus  
their conclusions are intact. I would like to thank Eric Braaten 
for providing me with this  
information.} in Eq.(22) of Ref.\cite{BPb}. 

It is once again emphasized that our approach for the calculation of  
the damping rate for the transverse gluons 
is quite different from the one taken by Braaten and Pisarski. 
We first constructed the gauge-independent resummed gluon 
self-energy, and then extracted the transverse function and
evaluated the discontinuity in the transverse function at 
$k_0=m_g$ and ${\bf k}=0$. By construction, the gauge-independence 
of our result is manifest.

\bigskip
\section{Summary and Discussion}
\smallskip


In this paper we have applied the $S$-matrix PT to hot QCD and calculated 
the effective gluon self-energy at one-loop order. We have found that 
the effective gluon self-energy, which is the sum of 
the resummed gluon self-energy and the resummed pinch contributions, is 
not only gauge-independent but also satisfies the transversality relation. 
Using this gauge-independent resummed gluon self-energy, we have calculated 
the damping rate for transverse gluon in the leading order and have shown 
that the result coincides with the one obtained by Braaten and Pisarski.

It was once pointed out by Baier, Kunstatter and Schiff~\cite{Kunstattera} that 
in covariant gauges (i.e., $ a_{COV}=1,\  b_{COV} =0$) 
there appear mass-shell singularities in the terms proportional to 
the gauge-fixing parameter $\xi$. In fact,  they showed 
explicitly that the imaginary part of the following integral
\be
   {\rm Im} \int dP \frac{1}{2}P^{\mu\nu}\bigl\{ I^{BB}_{\mu\nu} + I^{AB}_{\mu\nu}+
       I^{BS}_{\mu\nu}+ J^{B}_{\mu\nu} \bigr\}\Big\vert _{a=1, b=0}
\ee
develops poles on the physical mass shell and gives 
gauge-dependent gluon damping rate, unless an 
infrared regulator is maintained throughout the calculation. 
Note that the  gauge parameter $\xi$ appears only in the $B$-related terms.
(This issue also appeared in the calculation of the quark damping rate 
in the leading order~\cite{Kunstatterb}). 
We now know that the $B$-related term in the resummed gluon propagator 
in covariant gauges also gives rise to pinch contributions, and that this subtlety 
on mass-shell singularities in the covariant gauges 
dissappears once the pinch contributions are 
added to the resummed gluon self-energy.

We considered the one-loop quark-quark scattering amplitude to extract 
the resummed pinch contributions. Then 
there may arise an argument on the process-dependence of our result 
for the gauge-independent resummed gluon self-energy. Now let us inspect 
closely the expressions of the resummed pinch contributions 
given in Eqs.(\ref{IPB})-(\ref{IPST}). They are all made up of the 
terms which have at least one of the following factors:
\be
  {^{*}\Gamma_{\mu\nu}}(K), \quad {^{*}\Gamma_{\mu\alpha}}(K), 
   \quad {^{*}\Gamma_{\nu\alpha}}(K), \quad K^2_L \widetilde n_{\mu}(K), 
   \quad K^2_L \widetilde n_{\nu}(K)~.
\label{Factor}
\ee 
Those factors have come out when we used the identity given in Eq.(\ref{Identity}) 
and took away the resummed gluon propagator part, ${^{*}D^{\alpha\mu}}(K)$ 
or ${^{*}D^{\nu \beta}}(K)$. 
This fact reminds us of an idea of the {\it intrinsic} PT~\cite{rCP}. 
Since the gauge-independent 
resummed gluon self-energy has been obtained by adding the pinch contributions 
to the one-loop resummed gluon self energy ${^{*}\Pi}_{\mu\nu}(K)$ given in 
Eq.(\ref{PiSum}), there should be such terms in ${^{*}\Pi}_{\mu\nu}(K)$ that 
cancel against the pinch contributions. Those terms necessarily possess 
at least one of those factors given in Eq.(\ref{Factor}). 
Now the intrinsic PT algorithm tells us to start with    
three one-loop diagrams for the ${^{*}\Pi}_{\mu\nu}(K)$ depicted in Fig.2.(a)-(c), 
then to apply the Ward-Takahashi identities Eqs.(\ref{WTThree})-(\ref{WTFour}) 
to the corrected three- or four-point vertices and to throw out 
the factors given in Eq.(\ref{Factor}) which appeared as a result of  
the Ward-Takahashi identities. 
This intrinsic PT algorithm works fine for the case of the resummed gluon 
self-energy~\cite{Sasakic} and we can reach the same expression 
given in Eq.(\ref{InvPi}). Although Eq.(\ref{InvPi}) still 
contain  terms with such factors as given in Eq.(\ref{Factor}),  
these terms exactly cancel against the corresponding ones which 
come out of the products $~{^{*}\Gamma_{\lambda\mu}}^{\alpha}(P,K,Q)~ 
{^{*}\Gamma_{\alpha\nu}}^{\lambda}(Q,K,P)$ and 
$~{^{*}\Gamma_{0 \mu \alpha }}(P,K,Q)~
              {^{*}\Gamma^{\alpha}}_{\nu 0}(Q,K,P)$. 
The intrinsic PT do not resort to the consideration of any specific 
scattering process. And it succeeds to give the same expression 
that was obtained by the $S$-matrix PT in which we examined 
the one-loop quark-quark scattering. 
This supports implicitly the notion that our PT result on   
the resummed gluon self-energy is process independent. 

Furthermore the intrinsic PT can be applied to obtain the resummed gluon 
three-point function ${^{*}\Pi_{\mu \nu \lambda}}(P,Q,R)$. 
Then it can be shown that this effective three-point function 
${^{*}\Pi_{\mu \nu \lambda}}(P,Q,R)$ is {\it gauge independent} and also 
satisfies the {\it tree-level} Ward-Takahashi identity~\cite{Sasakic}
\be
     R^{\lambda}{^{*}\widehat{\Pi}_{\mu \nu \lambda}}(P,Q,R) = 
      {^{*}\widehat{\Pi}_{\mu\nu}}(P)-{^{*}\widehat{\Pi}_{\mu\nu}}(Q)~.
\ee
This means that the wave-function renormalization for the 
resummed gluon self-energy given in Eq.(\ref{InvPi}) contains the running of the 
QCD coupling and that, using its expression, we can obtain the correct 
thermal $\beta$ function in the leading order~\cite{Sasakid}.

\bigskip

\vspace{2cm}
\begin{center}
{\large\bf Acknowledgments}
\end{center}
\bigskip
\noindent
I would like to thank Alberto Sirlin for the hospitality extended to me 
at New York University and Rob Pisarski 
for the hospitality extended to me at Brookhaven National Laboratory.  
Part of this work was done at both places.   
This work is supported by the Grant-in-Aid for Scientific 
Research ((A)(1)(No 08304024)) of the Ministry of Education, 
Science and Culture of Japan.


\newpage
\appendix

\section{Pinch Contributions}
\smallskip

\noindent
(i) The contribution of the vertices of the first kind:
\be
  {^{*}\Pi^{P(V_1)}_{\mu\nu}}(K)= - Ng^2~ {^{*}\Gamma_{\mu\nu}}(K)  \int dP 
                     B(P)\frac{1}{P^2}    
\ee

\noindent
(ii) The contribution of the vertices of the second kind: 
\bea
 & & {^{*}\Pi^{P(V_2)}_{\mu\nu}}(K)= Ng^2 \int dP  \nonumber \\
  & & \qquad \times \Biggl\{\ \  - A(P)A(Q) \Biggl[~2~{^{*}\Gamma_{\mu\nu}}(K)~ 
+ \biggl\{ {^{*}\Gamma_{\mu\alpha}}(K)~ {V^{\alpha}}_{\nu}(P,K) 
  + (\mu \leftrightarrow \nu) \biggr\} \Biggr] \nonumber \\
& &\qquad \qquad -B(P)B(Q)\frac{1}{P^2 Q^2} P^{\lambda}P^{\tau}~
          {^{*}\Gamma_{\mu\lambda}}(K)~{^{*}\Gamma_{\nu\tau}}(K) \nonumber \\ 
& &\qquad \qquad + S(P)S(Q) \Biggl[\frac{p_0}{P^2}
       \biggl\{ K^2_L \widetilde n_{\mu}(K) \widetilde n^{\alpha}(Q)~ 
          {^{*}\Gamma_{\alpha\nu 0}}(Q,K,P)
              + (\mu \leftrightarrow \nu)\biggr\} \nonumber \\
& &\qquad \quad \qquad \qquad \qquad \qquad - \frac{1}{2} 
    \frac{p_0 q_0}{P^2 Q^2} \biggl\{ P^{\lambda}~ 
   {^{*}\Gamma_{\lambda\mu}}(K)~{^{*}\Gamma_{0 \nu 0}}(Q,K,P)
 + (\mu \leftrightarrow \nu)\biggr\}
              \Biggr] \nonumber \\
 & &\qquad \qquad + T(P)T(Q) \Biggl[ \biggl\{ 
   K^2_L \widetilde n_{\mu}(K) \bigl[K^2_L \widetilde n_{\nu}(K)-
           P^2_L \widetilde n_{\nu}(P) \bigr] + (\mu \leftrightarrow \nu) 
   \biggr\}  \nonumber \\
 & &\qquad \qquad \qquad \qquad \qquad \qquad + \biggl\{
   -\frac{1}{2}
       P^{\lambda}~{^{*}\Gamma_{\lambda\mu}}(K)~ 
         {^{*}\Gamma_{0 \nu 0}}(Q,K,P) + (\mu \leftrightarrow \nu)  \biggr\}
    \Biggr] \nonumber \\
 & &\qquad \qquad - A(Q)B(P)\frac{1}{P^2}
        \Biggl[ {^{*}\Gamma_{\mu\alpha}}(K) 
     \bigl[~{^{*}\Gamma^{\alpha}}_{\nu}(K)-{^{*}\Gamma^{\alpha}}_{\nu}(Q) \bigr]
           + (\mu \leftrightarrow \nu)   \Biggr]  \nonumber  \\
 & &\qquad \qquad + A(Q)S(P) \Biggl[ \biggl\{ K^2_L \widetilde n_{\mu}(K)  
 \widetilde n^{\alpha}(P) \bigl[ g_{\alpha\nu}+V_{\alpha\nu}(Q,K) \bigr] 
+ (\mu \leftrightarrow \nu) \biggr\} \nonumber  \\
 & &\qquad \qquad \qquad \qquad \qquad - \frac{p_0}{P^2} \biggl\{ 
  ~{^{*}\Gamma_{\mu}}^{\alpha}(K)~{^{*}\Gamma_{\alpha\nu 0}}(Q,K,P)  
   + (\mu \leftrightarrow \nu) \biggr\}  
 \Biggr] \nonumber \\
 & &\qquad \qquad + A(Q)T(P)\Biggl[ \biggl\{{^{*}\Gamma_{\mu\alpha}}(K)~ 
       {^{*}\Gamma^{\alpha}}_{\nu 0}(Q,K,P) + (\mu \leftrightarrow \nu) 
            \biggr\} \nonumber \\
 & &\qquad \qquad \qquad + 
    \biggl\{ \Bigl[ (-Q^2_T+ Q^2_L) 
     \frac{q_0}{Q^2\widetilde n^2(Q)} \widetilde n_{\mu}(Q)
      +  Q^2_T \frac{ Q_{\mu}}{Q^2}  \Bigr] 
    K^2_L  \widetilde n_{\nu}(K) + (\mu \leftrightarrow \nu) 
         \biggr\}   \Biggr]  \nonumber  \\ 
 & &\qquad \qquad - B(P)S(Q)\frac{1}{P^2} \Biggl[
     \biggl\{ \Bigl[K^2_L \widetilde n_{\mu}(K) + \frac{q_0}{Q^2}
  Q^{\lambda}{^{*}\Gamma_{\lambda\mu}}(K) \Bigr]
        \Bigl[K^2_L \widetilde n_{\nu}(K)-Q^2_L \widetilde n_{\nu}(Q) \Bigr] 
             \nonumber \\
 & &\qquad \qquad \qquad \qquad + (\mu \leftrightarrow \nu)  \biggr\} 
       + \frac{q_0}{Q^2}\biggl\{  
  K^2_L \widetilde n_{\mu}(K) Q^{\lambda}{^{*}\Gamma_{\lambda\nu}}(K)
    + (\mu \leftrightarrow \nu)   \biggr\} \Biggr]\nonumber \\ 
 & &\qquad \qquad + B(P)T(Q)\frac{1}{P^2} \Biggl[ 
   \biggl\{  K^2_L \widetilde n_{\mu}(K)
      Q^{\lambda}~{^{*}\Gamma_{\lambda\nu}}(K) 
                + (\mu \leftrightarrow \nu) \biggr\} \nonumber \\
 & &\qquad \qquad \qquad \qquad \qquad 
       + \biggl\{ Q^{\lambda}~{^{*}\Gamma_{\lambda\mu}}(K) 
     \Bigl[ K^2_L \widetilde n_{\nu}(K) - Q^2_L \widetilde n_{\nu}(Q) \Bigr] 
    + (\mu \leftrightarrow \nu)  \biggr\}  \Biggr] \nonumber \\
 & &\qquad \qquad  - S(Q)T(P) \Biggl[ \biggl\{ 
   \Bigl[K^2_L \widetilde n_{\mu}(K) + \frac{q_0}{Q^2}
  Q^{\lambda}{^{*}\Gamma_{\lambda\mu}}(K) \Bigr]~{^{*}\Gamma_{0 \nu 0}}(Q,K,P)
          + (\mu \leftrightarrow \nu)  \biggr\} \nonumber  \\
 & &\qquad \qquad \qquad \qquad \qquad  + \frac{q_0}{Q^2}\biggl\{  
  K^2_L \widetilde n_{\mu}(K) 
     \Bigl[2 K^2_L \widetilde n_{\nu}(K) - P^2_L \widetilde n_{\nu}(P) -
   Q^2_L \widetilde n_{\nu}(Q) \Bigr]  \nonumber  \\
 & &\qquad \qquad \qquad \qquad \qquad \qquad 
\qquad \qquad \qquad \qquad \qquad \qquad + (\mu \leftrightarrow \nu)  
\biggr\} \Biggr]   \Biggr\}
\eea

\noindent
(iii) The box contribution: 
\bea
  & & {^{*}\Pi^{P(Box)}_{\mu\nu}}(K)= Ng^2 \int dP  \nonumber \\
  & & \qquad \times \Biggl[~ 
        \frac{1}{2} B(P)B(Q)\frac{1}{P^2 Q^2} P^{\lambda}P^{\tau}~
          {^{*}\Gamma_{\mu\lambda}}(K)~{^{*}\Gamma_{\nu\tau}}(K) \nonumber \\
  & &\qquad \qquad - S(P)S(Q) \frac{p_0 q_0}{P^2 Q^2} 
       K^4_L \widetilde n_{\mu}(K) 
                          \widetilde n_{\nu}(K) \nonumber \\
  & &\qquad \qquad -T(P)T(Q) K^4_L \widetilde n_{\mu}(K) 
                          \widetilde n_{\nu}(K) \nonumber \\
  & &\qquad \qquad + A(Q)B(P)\frac{1}{P^2}~ 
         {^{*}\Gamma_{\mu\alpha}}(K)~{^{*}\Gamma^{\alpha}}_{\nu}(K)  
                      \nonumber  \\
 & &\qquad \qquad + B(P)S(Q)\frac{1}{P^2}
   \biggl[K^4_L \widetilde n_{\mu}(K) \widetilde n_{\nu}(K)   \nonumber  \\
   & &\qquad \qquad  \qquad \qquad - \frac{q_0}{Q^2} \Bigl\{ 
       P^{\lambda}~{^{*}\Gamma_{\lambda\mu}}(K) K^2_L  
      \widetilde n_{\nu}(K) + (\mu \leftrightarrow \nu)\Bigr\}  
                     \biggr]  \nonumber  \\
  & &\qquad \qquad + B(P)T(Q) \frac{1}{P^2} 
        \Bigl\{ P^{\lambda}~{^{*}\Gamma_{\lambda\mu}}(K) 
                   K^2_L \widetilde n_{\nu}(K) + (\mu \leftrightarrow \nu)
        \Bigr\}  \nonumber  \\
  & &\qquad \qquad + S(Q)T(P) \frac{2q_0}{Q^2} 
        K^4_L \widetilde n_{\mu}(K) \widetilde n_{\nu}(K) 
    \Biggr]
\eea

\section{Hard thermal loop for three-gluon vertex}

\noindent
The hard thermal loop for the three-gulon amplitude is expressed 
as~\cite{Taylor} 
\be
  \delta \Gamma_{\mu\nu\lambda}(P,Q,K) 
    = - \biggl[W_{\mu\nu\lambda}(P,K) - W_{\mu\nu\lambda}(Q,K) \biggr]~.  
\ee
The function $W_{\mu\nu\lambda}(P,K)$ is given by  
\be
     W_{\mu\nu\lambda}(P,K)=\frac{3 m_g^2}{4\pi}\int d\Omega 
        \widehat U_{\mu}\widehat U_{\nu}\widehat U_{\lambda}
               \frac{p_0}{\bigl[P \cdot \widehat U \bigr]
                 \bigl[K \cdot \widehat U \bigr]} 
\label{WWW}
\ee
where $m_g^2 =\frac{1}{9}Ng^2T^2$ is the gluon mass induced by the 
thermal medium and 
$\widehat U=(1, \widehat{\bf U})$ and  $\widehat{\bf U}$ is
a three-dimensional unit vector.
It is noted that $W_{\mu\nu\lambda}(P,K)$ is totally symmetric in 
indices $\mu$, $\nu$, and $\lambda$ and satisfies the following 
identities:
\bea
    g^{\mu\nu}W_{\mu\nu\lambda}(P,K) &=&0  
\label{B3}   \\
   K^{\lambda}W_{\mu\nu\lambda}(P,K) &=&-\delta \Pi_{\mu\nu}(P) 
      + 3 m_g^2 n_{\mu}n_{\nu}   
\label{B4} \\
   P^{\lambda}W_{\mu\nu\lambda}(P,K) &=&\frac{p_0}{k_0}
 \bigl[-\delta \Pi_{\mu\nu}(K)+ 3 m_g^2 n_{\mu}n_{\nu} \bigr]
\label{B5}
\eea
where 
\bea
        \delta \Pi_{\mu\nu}(K)&=&-\frac{3 m_g^2}{4\pi}\int d\Omega 
        \widehat U_{\mu}\widehat U_{\nu}
               \frac{k_0}{\bigl[K \cdot \widehat U \bigr]}  
     +  3 m_g^2 n_{\mu}n_{\nu}  \nonumber \\
         &=&\delta \Pi_T (K) P_{\mu\nu}(K) +\delta \Pi_L (K) Q_{\mu\nu}(K) 
\label{B6}
\eea
with~\cite{Silin}
\bea   
      \delta \Pi_T (K) &=& 3m^2_g \biggl\{-\frac{1}{4}\frac{k_0}{k} 
      \biggl[ \Bigl(\frac{k_0}{k}\Bigr)^2 -1 \biggr] {\rm ln}\frac{k_0+k}{k_0-k} 
       + \frac{1}{2}\Bigl(\frac{k_0}{k}\Bigr)^2   \biggr\}  \\
      \delta \Pi_L (K) &=& 3m^2_g \frac{K^2}{k^2} \biggl\{ \frac{1}{2} 
         \frac{k_0}{k}{\rm ln}\frac{k_0+k}{k_0-k} -1  \biggr\} \nonumber \\
        &=& -2 \delta \Pi_T (K) + 3m^2_g~.
\eea 
and $k=\vert {\bf k}\vert$.

The function $W_{\mu\nu\lambda}(P,K)$ may be expanded, in general, as
\bea 
     W_{\mu\nu\lambda}(P,K)&=& \alpha \bigl[n_\mu n_\nu n_\lambda \bigr] 
           \nonumber \\
        &+& \beta \bigl[ n_\mu n_\nu P_\lambda+n_\nu n_\lambda P_\mu 
       + n_\lambda n_\mu P_\nu \bigr] 
   + \beta'  \bigl[ n_\mu n_\nu K_\lambda+n_\nu n_\lambda K_\mu 
       + n_\lambda n_\mu K_\nu \bigr]   \nonumber \\
   &+& \delta \bigl[ n_\mu P_\nu P_\lambda+n_\nu P_\lambda P_\mu 
       + n_\lambda P_\mu P_\nu \bigr] 
   + \delta'  \bigl[ n_\mu K_\nu K_\lambda+n_\nu K_\lambda K_\mu 
       + n_\lambda K_\mu K_\nu \bigr]   \nonumber \\
  &+& \zeta \bigl[n_\mu P_\nu K_\lambda + n_\mu K_\nu P_\lambda + 
 n_\nu P_\lambda K_\mu + n_\nu K_\lambda P_\mu
       + n_\lambda P_\mu K_\nu + n_\lambda K_\mu P_\nu   \bigr] 
  \nonumber \\
  &+& \eta \bigl[P_\mu P_\nu P_\lambda \bigr]  + 
      \theta \bigl[K_\mu K_\nu K_\lambda \bigr] \nonumber \\
 &+& \kappa  \bigl[ P_\mu P_\nu K_\lambda+P_\nu P_\lambda K_\mu 
       + P_\lambda P_\mu K_\nu \bigr] + 
     \xi \bigl[ P_\mu K_\nu K_\lambda+P_\nu K_\lambda K_\mu 
       + P_\lambda K_\mu K_\nu \bigr]   \nonumber \\ 
 &+& \psi \bigl[g_{\mu\nu} n_{\lambda} + g_{\nu\lambda} n_{\mu} +
                   g_{\lambda\mu} n_{\nu} \bigr]   \nonumber \\ 
 &+& \chi \bigl[g_{\mu\nu} P_{\lambda} + g_{\nu\lambda} P_{\mu} +
                   g_{\lambda\mu} P_{\nu} \bigr] + 
     \omega \bigl[g_{\mu\nu} K_{\lambda} + g_{\nu\lambda} K_{\mu} +
                   g_{\lambda\mu} K_{\nu} \bigr]. 
\label{Wexpand}
\eea
Picking up the terms which are proportinal to $P_{\mu}$ from 
$W_{\mu\nu\lambda}(P,K)$, we find 
\be 
   W_{\mu\nu\lambda}(P,K) = P_{\mu}V_{\nu\lambda}(P,K) + \cdots 
\label{WexpandV}
\ee
with 
\bea
  V_{\nu\lambda}(P,K) &=& \beta(P,K) \bigl[ n_\nu n_\lambda \bigr] + 
   \delta(P,K) \bigl[n_\nu P_\lambda +  P_\nu n_\lambda \bigr]  
      \nonumber  \\
   &+& \zeta(P,K) \bigl[n_\nu K_\lambda +  K_\nu n_\lambda \bigr] 
      + \eta(P,K) \bigl[ P_\nu P_\lambda \bigr] +
     \kappa(P,K) \bigl[P_\nu K_\lambda +  K_\nu P_\lambda \bigr]  
   \nonumber \\  
   &+& \xi(P,K)\bigl[ K_\nu K_\lambda \bigr] +  \chi(P,K) g_{\nu\lambda}.
\label{Vexpand}
\eea
Obviously the function $V_{\nu\lambda}(P,K)$ is symmetric 
in indices $\nu$ and $\lambda$.

The identities (\ref{B3})-(\ref{B5}) provide many relations 
satisfied by 
the functions $\beta$, $\delta$, $\cdots$  in $V_{\nu\lambda}(P,K)$. 
In particular, (\ref{B4}) gives   
\be
   K^{\lambda} V_{\nu\lambda}(P,K)=\delta \Pi_T (P) \bigl[\frac{P_{\nu}}{P^2} + 
  \frac{p_0}{{\bf p}^2}\widetilde n_{\nu}(P) \bigr] 
      -\delta \Pi_L (P) \frac{p_0}{{\bf p}^2}\widetilde n_{\nu}(P)~,
\label{IdentityV}
\ee
which leads to the following relations: 
\bea
  & & k_0\beta + (K\cdot P)\delta  + K^2\zeta = 
    \bigl[ \delta \Pi_T (P) - \delta \Pi_L (P) \bigr] \frac{p_0}{{\bf p}^2}   
\label{WIbeta}   \\
 & & k_0\delta + (K\cdot P)\eta + K^2\kappa = 
      - \delta \Pi_T (P) \frac{1}{{\bf p}^2}  
+ \delta \Pi_L (P) \frac{p^2_0}{{\bf p}^2} \frac{1}{P^2}  
\label{WIdelta}   \\
  & & k_0\zeta + (K\cdot P)\kappa + K^2\xi + \chi =0~.  
\label{WIzeta}  
\eea

It is a formidable task to find out, in general, the expressions 
of the functions $\beta$, $\delta$, $\zeta$, $\eta$, $\kappa$,  
$\xi$ and $\chi$ in $V_{\nu\lambda}(P,K)$.  
However, when we restrict ourself to 
the special cases such as $\bf k = 0$ or $k_0=0$, it is not so hard to
obtain the relevant expressions.


\bigskip
\section{Effective three- and four-gluon vertices at ${\bf k}=0$ and 
              $k_0 = m_g$ }
\smallskip


Evaluating Eq.(\ref{WWW}) at ${\bf k}=0$, we find 
\bea
 W_{00i}(P,K) \Big\vert _{{\bf k}=0} &=& \frac{p_i p_0}{k_0P^2}
   \biggl[\frac{2}{3}\frac{{\bf p}^2}{P^2}X(P)+m^2_g  \biggr]  \\
 W_{0ji}(P,K) \Big\vert _{{\bf k}=0} &=& \frac{p_j p_i}{k_0P^2}
   \biggl[\Bigl(1+\frac{2}{3}\frac{{\bf p}^2}{P^2}\Bigr) X(P)+m^2_g \biggr]  
  \nonumber  \\
    & &+\frac{\delta_{ji}}{k_0} 
   \biggl[-\frac{1}{3}\frac{{\bf p}^2}{P^2}X(P)+m^2_g  \biggr]  \\
 W_{lji}(P,K) \Big\vert _{{\bf k}=0} &=& 
  \frac{p_l p_j p_i p_0}{k_0P^2{\bf p}^2}
\biggl[\Bigl(\frac{5}{3}+\frac{2}{3}\frac{{\bf p}^2}{P^2}\Bigr) X(P)+m^2_g 
           \biggr]  \nonumber \\
  & &- \Bigl[p_l \delta_{ji}+ p_j \delta_{il}+p_i \delta_{lj} \Bigr]
    \frac{p_0}{3k_0P^2} X(P)~,
\eea
where
\be
   X(P)=\frac{P^2}{{\bf p}^2}\bigl\{\delta \Pi_L (P) - 
           \delta \Pi_T (P) \bigr\}~.     
\ee 
With these informations we obtain 
\bea 
 & &{^{*}\Gamma_{00i}}(P,Q,K)~{^{*}\Gamma_{00i}}(P,Q,K) 
      \Big\vert _{{\bf k}=0, k_0 = m_g} \nonumber  \\
 & &\qquad \qquad =\frac{{\bf q}^2}{m_g^2 } 
     \biggl\{ 3 m_g^2 \biggl[ \frac{p_0}{P^2} + \frac{q_0}{Q^2}\biggr] + 
       2 \biggl[ \frac{p_0}{P^2} P_T^2 + \frac{q_0}{Q^2} Q_T^2  \biggr] \biggr\}^2 
          \\
 & &\qquad \qquad = \frac{{\bf q}^2}{m_g^2 }
    \biggl\{ 3 \biggl[ p_0 + \frac{q_0}{Q^2} m_g^2\biggr] - 
       \frac{p_0}{P^2} P_L^2 + 2\frac{q_0}{Q^2} Q_T^2  \biggr\}^2 
           \\     
 & &\qquad \qquad = {\bf q}^2 
     \biggl\{ 3 + \frac{1}{k_0}
       \biggl[ \frac{p_0}{P^2} P_L^2 + \frac{q_0}{Q^2} Q_L^2 
       \biggr] \biggr\}^2 ~,
\eea

\bea 
 & &{^{*}\Gamma_{0ji}}(P,Q,K)~{^{*}\Gamma_{0ji}}(P,Q,K) 
      \Big\vert _{{\bf k}=0, k_0 = m_g} \nonumber  \\
 & &\qquad \qquad =2 \biggl\{ -3q_0 + \frac{1}{k_0} (P_T^2 - Q_T^2) \biggr\}^2 
                      \nonumber  \\
  & &\qquad \qquad \quad  + \biggl\{ 3p_0 + 3(p_0 - q_0) 
           \frac{m_g^2 {\bf p}^2}{P^2 Q^2} + \frac{2}{k_0}
     \biggl[ \frac{p_0}{P^2} P_T^2 - \frac{q_0}{Q^2} Q_T^2  \biggr] \biggr\}^2 
            \\ 
 & &\qquad \qquad =\frac{1}{2 m_g^2} \biggl\{ 3Q^2 - (P_L^2 + 2 Q_T^2) \biggr\}^2 
                      \nonumber  \\
   & &\qquad \qquad \quad  + \frac{1}{ m_g^2} \biggl\{ 
             3 \biggl[ \frac{q_0}{Q^2} m_g^2 + p_0 q_0 \biggr] + 
      \biggl[ \frac{p_0}{P^2} P_L^2 + 2\frac{q_0}{Q^2} Q_T^2 
  \biggr] \biggr\}^2~,
\eea

\bea 
 & &{^{*}\Gamma_{lji}}(P,Q,K)~{^{*}\Gamma_{lji}}(P,Q,K) 
      \Big\vert _{{\bf k}=0, k_0 = m_g} \nonumber  \\
 & &\qquad \qquad = \frac{1}{{\bf p}^2} 
 \biggl\{ \frac{2}{k_0} \biggl[ \frac{p_0^3}{P^2} P_T^2 
            + \frac{q_0^3}{Q^2} Q_T^2 \biggr]  - 6 p_0 q_0  + 
    3 \frac{m_g^2 {\bf p}^2}{P^2 Q^2} ({\bf p}^2 - p_0 q_0 ) \biggr\}^2 
                           \nonumber  \\
    & &\qquad \qquad \quad  + \frac{6}{{\bf p}^2}  
   \biggl\{ \frac{1}{k_0} ( p_0 P_T^2  + q_0 Q_T^2 ) - 3 p_0 q_0 
           - {\bf p}^2 \biggr\}^2 
          + 12 {\bf p}^2~.
\eea

Since the required hard thermal loop for the four-gluon vertex 
is given by 
\be
  \delta \Gamma_{ii00}(K,-K,P,-P) \Big\vert _{{\bf k}=0, k_0 = m_g}
    = -3\biggl[ \frac{p_0}{p}{\rm ln}\frac{p_0+p}{p_0-p} 
    -  \frac{q_0}{q}{\rm ln}\frac{q_0+q}{q_0-q} \biggr]~,
\ee
we obtain
\be 
 {^{*}\Gamma_{ii00}}(K,-K,P,-P) \Big\vert _{{\bf k}=0, k_0 = m_g}
     = -6 - 6{\bf p}^2 \biggl[ \frac{1}{P^2} -  \frac{1}{Q^2} \biggr] 
   - \frac{4{\bf p}^2}{m_g^2}\biggl[ \frac{P_T^2}{P^2} -  \frac{Q_T^2}{Q^2} \biggr]
\ee




\bigskip
\newpage
\noindent
{\large\bf Figure Caption}
\medskip

\noindent
Fig.1

\noindent
The ghost-gluon vertex in linear gauges which preserve 
rotational invariance. The wavy and dashed lines denote 
gluon and ghost fields, respectively.

\medskip

\noindent
Fig.2

\noindent
(a) The resummed gluon self-energy diagram with three-gluon interactions. \\
(b) The tadpole diagram for the resummed gluon self-energy. \\
(c) The ghost diagram for the resummed gluon self-energy. \\
The blobs in the gluon propagators and vertices represent that 
they are effective quantities with the hard thermal loop 
corrections included.
\medskip

\noindent
Fig.3

\noindent
The gluon self-energy diagram for the quark-quark scattering.
The solid lines represent quark fields.

\medskip

\noindent
Fig.4

\noindent
(a) The vertex diagrams of the first kind for 
the quark-quark scattering. \\
(b) Their pinch contribution.

\medskip

\noindent
Fig.5

\noindent
(a) The vertex diagram of the second kind 
for the quark-quark scattering. 
(b) Its pinch contribution.

\medskip

\noindent
Fig.6

\noindent
(a) The box diagrams for the quark-quark scattering.   
(b) Their pinch contribution.



\newenvironment{texdraw}{\leavevmode\btexdraw}{\etexdraw}

\def\setRevDate $#1 #2 #3${\def\TeXdrawId{TeXdraw V1R3 revised <#2>}}
\setRevDate $Date: Wed Mar  3 12:01:12 MST 1993$

\chardef\catamp=\the\catcode`\@
\catcode`\@=11
\ifx\TeXdraw@included\undefined\global\let\TeXdraw@included=\relax\else
\errhelp{TeXdraw needs to be input only once outside of any groups.}%
\errmessage{Multiple call to include TeXdraw ignored}%
\expandafter\endinput\fi
\long
\def\centertexdraw #1{\hbox to \hsize{\hss
\btexdraw #1\etexdraw
\hss}}
\def\btexdraw {\x@pix=0 \y@pix=0
\x@segoffpix=\x@pix  \y@segoffpix=\y@pix
\t@exdrawdef
\setbox\t@xdbox=\vbox\bgroup\offinterlineskip
\global\d@bs=0 
\t@extonlytrue 
\p@osinitfalse
\savemove \x@pix \y@pix 
\m@pendingfalse
\p@osinitfalse          
\p@athfalse}
\def\etexdraw {\ift@extonly \else
\t@drclose
\fi
\egroup 
\ifdim \wd\t@xdbox>0pt
\errmessage{TeXdraw box non-zero size, possible extraneous text}%
\fi
\maxhvpos 
\pixtodim \xminpix \l@lxpos  \pixtodim \yminpix \l@lypos
\pixtobp {-\xminpix}\l@lxbp  \pixtobp {-\yminpix}\l@lybp
\vbox {\offinterlineskip
\ift@extonly \else
\includepsfile{\p@sfile}{\the\l@lxbp}{\the\l@lybp}%
{\the\hdrawsize}{\the\vdrawsize}%
\fi
\vskip\vdrawsize
\vskip \l@lypos
\hbox {\hskip -\l@lxpos
\box\t@xdbox
\hskip \hdrawsize
\hskip \l@lxpos}%
\vskip -\l@lypos\relax}}
\def\drawdim #1 {\def\d@dim{#1\relax}}
\def\setunitscale #1 {\edef\u@nitsc{#1}%
\realmult \u@nitsc  \s@egsc \d@sc}
\def\relunitscale #1 {\realmult {#1}\u@nitsc \u@nitsc
\realmult \u@nitsc \s@egsc \d@sc}
\def\setsegscale #1 {\edef\s@egsc {#1}%
\realmult \u@nitsc \s@egsc \d@sc}
\def\relsegscale #1 {\realmult {#1}\s@egsc \s@egsc
\realmult \u@nitsc \s@egsc \d@sc}
\def\bsegment {\ifp@ath
\flushmove
\fi
\begingroup
\x@segoffpix=\x@pix
\y@segoffpix=\y@pix
\setsegscale 1
\global\advance \d@bs by 1 }
\def\esegment {\endgroup
\ifnum \d@bs=0
\writetx {es}%
\else
\global\advance \d@bs by -1
\fi}
\def\savecurrpos (#1 #2){\getsympos (#1 #2)\a@rgx\a@rgy
\s@etcsn \a@rgx {\the\x@pix}%
\s@etcsn \a@rgy {\the\y@pix}}%
\def\savepos (#1 #2)(#3 #4){\getpos (#1 #2)\a@rgx\a@rgy
\coordtopix \a@rgx \t@pixa
\advance \t@pixa by \x@segoffpix
\coordtopix \a@rgy \t@pixb
\advance \t@pixb by \y@segoffpix
\getsympos (#3 #4)\a@rgx\a@rgy
\s@etcsn \a@rgx {\the\t@pixa}%
\s@etcsn \a@rgy {\the\t@pixb}}
\def\linewd #1 {\coordtopix {#1}\t@pixa
\flushbs
\writetx {\the\t@pixa\space sl}}
\def\setgray #1 {\flushbs
\writetx {#1 sg}}
\def\lpatt (#1){\listtopix (#1)\p@ixlist
\flushbs
\writetx {[\p@ixlist] sd}}
\def\lvec (#1 #2){\getpos (#1 #2)\a@rgx\a@rgy
\s@etpospix \a@rgx \a@rgy
\writeps {\the\x@pix\space \the\y@pix\space lv}}
\def\rlvec (#1 #2){\getpos (#1 #2)\a@rgx\a@rgy
\r@elpospix \a@rgx \a@rgy
\writeps {\the\x@pix\space \the\y@pix\space lv}}
\def\move (#1 #2){\getpos (#1 #2)\a@rgx\a@rgy
\s@etpospix \a@rgx \a@rgy
\savemove \x@pix \y@pix}
\def\rmove (#1 #2){\getpos (#1 #2)\a@rgx\a@rgy
\r@elpospix \a@rgx \a@rgy
\savemove \x@pix \y@pix}
\def\lcir r:#1 {\coordtopix {#1}\t@pixa
\writeps {\the\t@pixa\space cr}%
\r@elupd \t@pixa \t@pixa
\r@elupd {-\t@pixa}{-\t@pixa}}
\def\fcir f:#1 r:#2 {\coordtopix {#2}\t@pixa
\writeps {#1 \the\t@pixa\space fc}%
\r@elupd \t@pixa \t@pixa
\r@elupd {-\t@pixa}{-\t@pixa}}
\def\lellip rx:#1 ry:#2 {\coordtopix {#1}\t@pixa
\coordtopix {#2}\t@pixb
\writeps {\the\t@pixa\space \the\t@pixb\space el}%
\r@elupd \t@pixa \t@pixb
\r@elupd {-\t@pixa}{-\t@pixb}}
\def\larc r:#1 sd:#2 ed:#3 {\coordtopix {#1}\t@pixa
\writeps {\the\t@pixa\space #2 #3 ar}}
\def\ifill f:#1 {\writeps {#1 fl}}
\def\lfill f:#1 {\writeps {#1 fp}}
\def\htext #1{\def\testit {#1}%
\ifx \testit\l@paren
\let\next=\h@move
\else
\let\next=\h@text
\fi
\next{#1}}
\def\rtext td:#1 #2{\def\testit {#2}%
\ifx \testit\l@paren
\let\next=\r@move
\else
\let\next=\r@text
\fi
\next td:#1 {#2}}
\def\vtext {\rtext td:90 }
\def\textref h:#1 v:#2 {\ifx #1R%
\edef\l@stuff {\hss}\edef\r@stuff {}%
\else
\ifx #1C%
\edef\l@stuff {\hss}\edef\r@stuff {\hss}%
\else  
\edef\l@stuff {}\edef\r@stuff {\hss}%
\fi
\fi
\ifx #2T%
\edef\t@stuff {}\edef\b@stuff {\vss}%
\else
\ifx #2C%
\edef\t@stuff {\vss}\edef\b@stuff {\vss}%
\else  
\edef\t@stuff {\vss}\edef\b@stuff {}%
\fi
\fi}
\def\avec (#1 #2){\getpos (#1 #2)\a@rgx\a@rgy
\s@etpospix \a@rgx \a@rgy
\writeps {\the\x@pix\space \the\y@pix\space (\a@type)
\the\a@lenpix\space \the\a@widpix\space av}}
\def\ravec (#1 #2){\getpos (#1 #2)\a@rgx\a@rgy
\r@elpospix \a@rgx \a@rgy
\writeps {\the\x@pix\space \the\y@pix\space (\a@type)
\the\a@lenpix\space \the\a@widpix\space av}}
\def\arrowheadsize l:#1 w:#2 {\coordtopix{#1}\a@lenpix
\coordtopix{#2}\a@widpix}
\def\arrowheadtype t:#1 {\edef\a@type{#1}}
\def\curvytype#1{\def\curv@type{#1}}\curvytype{4}%
\def\curvyheight#1{\def\curv@height{#1}}\curvyheight{10}%
\def\curvylength#1{\def\curv@length{#1}}\curvylength{10}%
\def\drawcurvyphoton around (#1 #2) from (#3 #4) to (#5 #6)%
{\getpos (#1 #2)\a@rgx\a@rgy
\coordtopix \a@rgx \t@pixa \advance \t@pixa by \x@segoffpix
\coordtopix \a@rgy \t@pixb \advance \t@pixb by \y@segoffpix
\writeps {mark \the\t@pixa\space \the\t@pixb}%
\getpos (#3 #4)\a@rgx\a@rgy
\s@etpospix \a@rgx\a@rgy
\writeps {\the\x@pix\space \the\y@pix}%
\getpos (#5 #6)\a@rgx\a@rgy
\s@etpospix \a@rgx\a@rgy
\writeps {\the\x@pix\space \the\y@pix}%
\writeps {\curv@height\space \curv@length\space \curv@type\space%
curvyphoton}%
}%
\def\drawcurvygluon around (#1 #2) from (#3 #4) to (#5 #6)%
{\getpos (#1 #2)\a@rgx\a@rgy
\coordtopix \a@rgx \t@pixa \advance \t@pixa by \x@segoffpix
\coordtopix \a@rgy \t@pixb \advance \t@pixb by \y@segoffpix
\writeps {mark \the\t@pixa\space \the\t@pixb}%
\getpos (#3 #4)\a@rgx\a@rgy
\s@etpospix \a@rgx\a@rgy
\writeps {\the\x@pix\space \the\y@pix}%
\getpos (#5 #6)\a@rgx\a@rgy
\s@etpospix \a@rgx\a@rgy
\writeps {\the\x@pix\space \the\y@pix}%
\writeps {\curv@height\space 2 mul \curv@length\space \curv@type\space%
curvygluon}%
}%
\def\blobfreq#1{\def\bl@bfreq{#1}}\blobfreq{0.2}%
\def\blobangle#1{\def\bl@bangle{#1}}\blobangle{0}%
\def\hatchedblob#1{\def\bl@btype{(#1)}}\hatchedblob{B}%
\def\grayblob#1{\def\bl@btype{#1}}%
\def\drawblob xsize:#1 ysize:#2 at (#3 #4)%
{\getpos (#3 #4)\a@rgx\a@rgy \s@etpospix \a@rgx\a@rgy
\writeps{\the\x@pix\space \the\y@pix}%
\getpos (#1 #2)\a@rgx\a@rgy 
\coordtopix \a@rgx\t@pixa \coordtopix\a@rgy\t@pixb \writeps
{\the\t@pixa\space\the\t@pixb\space\bl@bangle\space\bl@bfreq\space\bl@btype}%
\writeps {blob}%
\rmove (-#1 -#2)\rmove (#1 #2)\rmove (#1 #2)\rmove (-#1 -#2)}%
\def\drawbb {\bsegment
\drawdim bp
\setunitscale 0.24
\linewd 1
\writeps {\the\xminpix\space \the\yminpix\space mv}%
\writeps {\the\xminpix\space \the\ymaxpix\space lv}%
\writeps {\the\xmaxpix\space \the\ymaxpix\space lv}%
\writeps {\the\xmaxpix\space \the\yminpix\space lv}%
\writeps {\the\xminpix\space \the\yminpix\space lv}%
\esegment}
\def\getpos (#1 #2)#3#4{\g@etargxy #1 #2 {} \\#3#4%
\c@heckast #3%
\ifa@st
\g@etsympix #3\t@pixa
\advance \t@pixa by -\x@segoffpix
\pixtocoord \t@pixa #3
\fi
\c@heckast #4%
\ifa@st
\g@etsympix #4\t@pixa
\advance \t@pixa by -\y@segoffpix
\pixtocoord \t@pixa #4
\fi}
\def\getsympos (#1 #2)#3#4{\g@etargxy #1 #2 {} \\#3#4%
\c@heckast #3%
\ifa@st \else
\errmessage {TeXdraw: invalid symbolic coordinate}
\fi
\c@heckast #4%
\ifa@st \else
\errmessage {TeXdraw: invalid symbolic coordinate}
\fi}
\def\listtopix (#1)#2{\def #2{}%
\edef\l@ist {#1 }%
\t@countc=0
\loop
\expandafter\g@etitem \l@ist \\\a@rgx\l@ist
\a@pppix \a@rgx #2
\ifx \l@ist\empty
\t@countc=1
\fi
\ifnum \t@countc=0
\repeat}
\def\realmult #1#2#3{\dimen0=#1pt
\dimen2=#2\dimen0
\edef #3{\expandafter\c@lean\the\dimen2}}
\def\intdiv #1#2#3{\t@counta=#1
\t@countb=#2
\ifnum \t@countb<0
\t@counta=-\t@counta
\t@countb=-\t@countb
\fi
\t@countd=1
\ifnum \t@counta<0
\t@counta=-\t@counta
\t@countd=-1
\fi
\t@countc=\t@counta  \divide \t@countc by \t@countb
\t@counte=\t@countc  \multiply \t@counte by \t@countb
\advance \t@counta by -\t@counte
\t@counte=-1
\loop
\advance \t@counte by 1
\ifnum \t@counte<16
\multiply \t@countc by 2 
\multiply \t@counta by 2 
\ifnum \t@counta<\t@countb \else
\advance \t@countc by 1     
\advance \t@counta by -\t@countb 
\fi
\repeat
\divide \t@countb by 2        
\ifnum \t@counta<\t@countb    
\advance \t@countc by 1
\fi
\ifnum \t@countd<0            
\t@countc=-\t@countc
\fi
\dimen0=\t@countc sp          
\edef #3{\expandafter\c@lean\the\dimen0}}
\outer\def\gnewif #1{\count@\escapechar \escapechar\m@ne
\expandafter\expandafter\expandafter
\edef\@if #1{true}{\global\let\noexpand#1=\noexpand\iftrue}%
\expandafter\expandafter\expandafter
\edef\@if #1{false}{\global\let\noexpand#1=\noexpand\iffalse}%
\@if#1{false}\escapechar\count@} 
\def\@if #1#2{\csname\expandafter\if@\string#1#2\endcsname}
{\uccode`1=`i \uccode`2=`f \uppercase{\gdef\if@12{}}} 
\def\coordtopix #1#2{\dimen0=#1\d@dim
\dimen2=\d@sc\dimen0
\t@counta=\dimen2    
\t@countb=\s@ppix
\divide \t@countb by 2
\ifnum \t@counta<0  
\advance \t@counta by -\t@countb
\else
\advance \t@counta by \t@countb
\fi
\divide \t@counta by \s@ppix
#2=\t@counta}
\def\pixtocoord #1#2{\t@counta=#1%
\multiply \t@counta by \s@ppix
\dimen0=\d@sc\d@dim
\t@countb=\dimen0
\intdiv \t@counta \t@countb #2}
\def\pixtodim #1#2{\t@countb=#1%
\multiply \t@countb by \s@ppix
#2=\t@countb sp\relax}
\def\pixtobp #1#2{\dimen0=\p@sfactor pt
\t@counta=\dimen0
\multiply \t@counta by #1%
\ifnum \t@counta < 0
\advance \t@counta by -32768
\else
\advance \t@counta by 32768
\fi
\divide \t@counta by 65536
#2=\t@counta}
\newcount\t@counta \newcount\t@countb
\newcount\t@countc \newcount\t@countd
\newcount\t@counte
\newcount\t@pixa \newcount\t@pixb
\newcount\t@pixc \newcount\t@pixd
\let\l@lxbp=\t@pixa \let\l@lybp=\t@pixb 
\let\u@rxbp=\t@pixc \let\u@rybp=\t@pixd
\newdimen\t@xpos \newdimen\t@ypos
\let\l@lxpos=\t@xpos \let\l@lypos=\t@ypos
\newcount\xminpix \newcount\xmaxpix
\newcount\yminpix \newcount\ymaxpix
\newcount\a@lenpix \newcount\a@widpix
\newcount\x@pix \newcount\y@pix
\newcount\x@segoffpix \newcount\y@segoffpix
\newcount\x@savepix \newcount\y@savepix
\newcount\s@ppix 
\newcount\d@bs
\newcount\t@xdnum
\global\t@xdnum=0
\newdimen\hdrawsize \newdimen\vdrawsize
\newbox\t@xdbox
\newwrite\drawfile
\newif\ifm@pending
\newif\ifp@ath
\newif\ifa@st
\gnewif \ift@extonly
\gnewif\ifp@osinit
\def\l@paren{(}
\def\a@st{*}
\catcode`\%=12
  \def\p@b {
\catcode`\%=14
\catcode`\{=12 \catcode`\}=12  \catcode`\u=1 \catcode`\v=2
  \def\l@br u{v \def\r@br u}v
\catcode `\{=1 \catcode`\}=2   \catcode`\u=11 \catcode`\v=11
{\catcode`\p=12 \catcode`\t=12
 \gdef\c@lean #1pt{#1}}
\def\sppix#1/#2 {\dimen0=1#2 \s@ppix=\dimen0
\t@counta=#1%
\divide \t@counta by 2
\advance \s@ppix by \t@counta
\divide \s@ppix by #1%
\t@counta=\s@ppix
\multiply \t@counta by 65536 
\advance \t@counta by 32891  
\divide \t@counta by 65782 
\dimen0=\t@counta sp
\edef\p@sfactor {\expandafter\c@lean\the\dimen0}}
\def\g@etargxy #1 #2 #3 #4\\#5#6{\def #5{#1}%
\ifx #5\empty
\g@etargxy #2 #3 #4 \\#5#6%
\else
\def #6{#2}%
\def\next {#3}%
\ifx \next\empty \else
\errmessage {TeXdraw: invalid coordinate}%
\fi
\fi}
\def\c@heckast #1{\expandafter
\c@heckastll #1\\}
\def\c@heckastll #1#2\\{\def\testit {#1}%
\ifx \testit\a@st
\a@sttrue
\else
\a@stfalse
\fi}
\def\g@etsympix #1#2{\expandafter
\ifx \csname #1\endcsname \relax
\errmessage {TeXdraw: undefined symbolic coordinate}%
\fi
#2=\csname #1\endcsname}
\def\s@etcsn #1#2{\expandafter
\xdef\csname#1\endcsname {#2}}
\def\g@etitem #1 #2\\#3#4{\edef #4{#2}\edef #3{#1}}
\def\a@pppix #1#2{\edef\next  {#1}%
\ifx \next\empty \else
\coordtopix {#1}\t@pixa
\ifx #2\empty
\edef #2{\the\t@pixa}%
\else
\edef #2{#2 \the\t@pixa}%
\fi
\fi}
\def\s@etpospix #1#2{\coordtopix {#1}\x@pix
\advance \x@pix by \x@segoffpix
\coordtopix {#2}\y@pix
\advance \y@pix by \y@segoffpix
\u@pdateminmax \x@pix \y@pix}
\def\r@elpospix #1#2{\coordtopix {#1}\t@pixa
\advance \x@pix by \t@pixa
\coordtopix {#2}\t@pixa
\advance \y@pix by \t@pixa
\u@pdateminmax \x@pix \y@pix}
\def\r@elupd #1#2{\t@counta=\x@pix
\advance\t@counta by #1%
\t@countb=\y@pix
\advance\t@countb by #2%
\u@pdateminmax \t@counta \t@countb}
\def\u@pdateminmax #1#2{\ifnum #1>\xmaxpix
\global\xmaxpix=#1%
\fi
\ifnum #1<\xminpix
\global\xminpix=#1%
\fi
\ifnum #2>\ymaxpix
\global\ymaxpix=#2%
\fi
\ifnum #2<\yminpix
\global\yminpix=#2%
\fi}
\def\maxhvpos {\t@pixa=\xmaxpix
\advance \t@pixa by -\xminpix
\pixtodim  \t@pixa {\dimen2}%
\global\hdrawsize=\dimen2
\t@pixa=\ymaxpix
\advance \t@pixa by -\yminpix
\pixtodim \t@pixa {\dimen2}%
\global\vdrawsize=\dimen2\relax}
\def\savemove #1#2{\x@savepix=#1\y@savepix=#2%
\m@pendingtrue
\ifp@osinit \else
\p@osinittrue
\global\xminpix=\x@savepix \global\yminpix=\y@savepix
\global\xmaxpix=\x@savepix \global\ymaxpix=\y@savepix
\fi}
\def\flushmove {\p@osinittrue
\ifm@pending
\writetx {\the\x@savepix\space \the\y@savepix\space mv}%
\m@pendingfalse
\p@athfalse
\fi}
\def\flushbs {\loop
\ifnum \d@bs>0
\writetx {bs}%
\global\advance \d@bs by -1
\repeat}
\def\h@move #1#2 #3)#4{\move (#2 #3)%
\h@text {#4}}
\def\h@text #1{\pixtodim \x@pix \t@xpos
\pixtodim \y@pix \t@ypos
\vbox to 0pt{\normalbaselines
\t@stuff
\kern -\t@ypos
\hbox to 0pt{\l@stuff
\kern \t@xpos
\hbox {#1}%
\kern -\t@xpos
\r@stuff}%
\kern \t@ypos
\b@stuff\relax}}
\def\r@move td:#1 #2#3 #4)#5{\move (#3 #4)%
\r@text td:#1 {#5}}
\def\r@text td:#1 #2{\pixtodim \x@pix \t@xpos
\pixtodim \y@pix \t@ypos
\vbox to 0pt{\kern -\t@ypos
\hbox to 0pt{\kern \t@xpos
\rottxt{#1}{#2}%
\hss}%
\vss}}
\def\rottxt #1#2{\rotsclTeX{#1}{1}{1}{\z@sb{#2}}}%
\def\z@sb #1{\vbox to 0pt{\normalbaselines
\t@stuff
\hbox to 0pt{\l@stuff
\hbox {#1}%
\r@stuff}%
\b@stuff}}
\def\t@exdrawdef {\sppix 300/in 
\drawdim in   
\edef\u@nitsc {1}%
\setsegscale 1    
\arrowheadsize l:0.16 w:0.08
\arrowheadtype t:T
\textref h:L v:B }
\def\writeps #1{\flushbs
\flushmove
\p@athtrue
\writetx {#1}}
\def\writetx #1{\ift@extonly
\t@extonlyfalse
\t@dropen
\fi
\w@rps {#1}}
\def\w@rps #1{\immediate\write\drawfile {#1}}
\def\t@dropen {%
\global\advance \t@xdnum by 1
\ifnum \t@xdnum<10
\xdef\p@sfile {\jobname.ps\the\t@xdnum}
\else
\xdef\p@sfile {\jobname.p\the\t@xdnum}
\fi
\immediate\openout\drawfile=\p@sfile
\w@rps {\p@b PS-Adobe-3.0 EPSF-3.0}%
\w@rps {\p@p BoundingBox: (atend)}%
\w@rps {\p@p Title: TeXdraw drawing: \p@sfile}%
\w@rps {\p@p Pages: 1 1}%
\w@rps {\p@p Creator: TeXdraw V1R3}%
\w@rps {\p@p CreationDate: \the\year/\the\month/\the\day}%
\w@rps {\p@p DocumentSuppliedResources: ProcSet TeXDraw 2.2 2}%
\w@rps {\p@p DocumentData: Clean7Bit}%
\w@rps {\p@p EndComments}%
\w@rps {\p@p BeginDefaults}%
\w@rps {\p@p PageNeededResources: ProcSet TeXDraw 2.2 2}%
\w@rps {\p@p EndDefaults}%
\w@rps {\p@p BeginProlog}%
\w@rps {\p@p BeginResource: ProcSet TeXDraw 2.2 2 14696 10668}%
\w@rps {\p@p VMlocation: local}%
\w@rps {\p@p VMusage: 14696 10668}%
\w@rps {/setglobal where}%
\w@rps {{pop currentglobal false setglobal} if}%
\w@rps {/setpacking where}%
\w@rps {{pop currentpacking false setpacking} if}%
\w@rps {29 dict dup begin}%
\w@rps {62 dict dup begin}%
\w@rps {/rad 0 def /radx 0 def /rady 0 def /svm matrix def}%
\w@rps {/hhwid 0 def /hlen 0 def /ah 0 def /tipy 0 def}%
\w@rps {/tipx 0 def /taily 0 def /tailx 0 def /dx 0 def}%
\w@rps {/dy 0 def /alen 0 def /blen 0 def}%
\w@rps {/i 0 def /y1 0 def /x1 0 def /y0 0 def /x0 0 def}%
\w@rps {/movetoNeeded 0 def}%
\w@rps {/y3 0 def /x3 0 def /y2 0 def /x2 0 def}%
\w@rps {/p1y 0 def /p1x 0 def /p2y 0 def /p2x 0 def}%
\w@rps {/p0y 0 def /p0x 0 def /p3y 0 def /p3x 0 def}%
\w@rps {/n 0 def /y 0 def /x 0 def}%
\w@rps {/anglefactor 0 def /elemlength 0 def /excursion 0 def}%
\w@rps {/endy 0 def /endx 0 def /beginy 0 def /beginx 0 def}%
\w@rps {/centery 0 def /centerx 0 def /startangle 0 def }%
\w@rps {/startradius 0 def /endradius 0 def /elemcount 0 def}%
\w@rps {/smallincrement 0 def /angleincrement 0 def /radiusincrement 0 def}%
\w@rps {/ifleft false def /ifright false def /iffill false def}%
\w@rps {/freq 1 def /angle 0 def /yrad 0 def /xrad 0 def /y 0 def /x 0 def}%
\w@rps {/saved 0 def}%
\w@rps {end}%
\w@rps {/dbdef {1 index exch 0 put 0 begin bind end def}}%
\w@rps {dup 3 4 index put dup 5 4 index put bind def pop}%
\w@rps {/bdef {bind def} bind def}%
\w@rps {/mv {stroke moveto} bdef}%
\w@rps {/lv {lineto} bdef}%
\w@rps {/st {currentpoint stroke moveto} bdef}%
\w@rps {/sl {st setlinewidth} bdef}%
\w@rps {/sd {st 0 setdash} bdef}%
\w@rps {/sg {st setgray} bdef}%
\w@rps {/bs {gsave} bdef /es {stroke grestore} bdef}%
\w@rps {/cv {curveto} bdef}%
\w@rps {/cr \l@br 0 begin}%
\w@rps {gsave /rad exch def currentpoint newpath rad 0 360 arc}%
\w@rps {stroke grestore end\r@br\space 0 dbdef}%
\w@rps {/fc \l@br 0 begin}%
\w@rps {gsave /rad exch def setgray currentpoint newpath}%
\w@rps {rad 0 360 arc fill grestore end\r@br\space 0 dbdef}%
\w@rps {/ar {gsave currentpoint newpath 5 2 roll arc stroke grestore} bdef}%
\w@rps {/el \l@br 0 begin gsave /rady exch def /radx exch def}%
\w@rps {svm currentmatrix currentpoint translate}%
\w@rps {radx rady scale newpath 0 0 1 0 360 arc}%
\w@rps {setmatrix stroke grestore end\r@br\space 0 dbdef}%
\w@rps {/fl \l@br gsave closepath setgray fill grestore}%
\w@rps {currentpoint newpath moveto\r@br\space bdef}%
\w@rps {/fp \l@br gsave closepath setgray fill grestore}%
\w@rps {currentpoint stroke moveto\r@br\space bdef}%
\w@rps {/av \l@br 0 begin /hhwid exch 2 div def /hlen exch def}%
\w@rps {/ah exch def /tipy exch def /tipx exch def}%
\w@rps {currentpoint /taily exch def /tailx exch def}%
\w@rps {/dx tipx tailx sub def /dy tipy taily sub def}%
\w@rps {/alen dx dx mul dy dy mul add sqrt def}%
\w@rps {/blen alen hlen sub def}%
\w@rps {gsave tailx taily translate dy dx atan rotate}%
\w@rps {(V) ah ne {blen 0 gt {blen 0 lineto} if} {alen 0 lineto} ifelse}%
\w@rps {stroke blen hhwid neg moveto alen 0 lineto blen hhwid lineto}%
\w@rps {(T) ah eq {closepath} if}%
\w@rps {(W) ah eq {gsave 1 setgray fill grestore closepath} if}%
\w@rps {(F) ah eq {fill} {stroke} ifelse}%
\w@rps {grestore tipx tipy moveto end\r@br\space 0 dbdef}%
\w@rps {/setupcurvy \l@br 0 begin}%
\w@rps {dup 0 eq {1 add} if /anglefactor exch def}%
\w@rps {abs dup 0 eq {1 add} if /elemlength exch def /excursion exch def}%
\w@rps {/endy exch def /endx exch def}%
\w@rps {/beginy exch def /beginx exch def}%
\w@rps {/centery exch def /centerx exch def}%
\w@rps {cleartomark}%
\w@rps {/startangle beginy centery sub beginx centerx sub atan def}%
\w@rps {/startradius beginy centery sub dup mul }%
\w@rps {beginx centerx sub dup mul add sqrt def}%
\w@rps {/endradius endy centery sub dup mul }%
\w@rps {endx centerx sub dup mul add sqrt def}%
\w@rps {endradius startradius sub }%
\w@rps {endy centery sub endx centerx sub atan }%
\w@rps {startangle 2 copy le {exch 360 add exch} if sub dup}%
\w@rps {elemlength startradius endradius add atan dup add}%
\w@rps {div round abs cvi dup 0 eq {1 add} if}%
\w@rps {dup /elemcount exch def }%
\w@rps {div dup anglefactor div dup /smallincrement exch def}%
\w@rps {sub /angleincrement exch def}%
\w@rps {elemcount div /radiusincrement exch def}%
\w@rps {gsave newpath}%
\w@rps {startangle dup cos startradius mul }%
\w@rps {centerx add exch }%
\w@rps {sin startradius mul centery add moveto}%
\w@rps {end \r@br 0 dbdef}%
\w@rps {/curvyphoton \l@br 0 begin}%
\w@rps {setupcurvy}%
\w@rps {elemcount \l@br /startangle startangle smallincrement add def}%
\w@rps {/startradius startradius excursion add def}%
\w@rps {startangle dup cos startradius mul }%
\w@rps {centerx add exch }%
\w@rps {sin startradius mul centery add}%
\w@rps {/excursion excursion neg def}%
\w@rps {/startangle startangle angleincrement add }%
\w@rps {smallincrement sub def}%
\w@rps {/startradius startradius radiusincrement add def}%
\w@rps {startangle dup cos startradius mul }%
\w@rps {centerx add exch }%
\w@rps {sin startradius mul centery add}%
\w@rps {/startradius startradius excursion add def}%
\w@rps {/startangle startangle smallincrement add def}%
\w@rps {startangle dup cos startradius mul }%
\w@rps {centerx add exch }%
\w@rps {sin startradius mul centery add curveto\r@br repeat}%
\w@rps {stroke grestore end}%
\w@rps {\r@br 0 dbdef}%
\w@rps {/curvygluon \l@br 0 begin}%
\w@rps {setupcurvy /radiusincrement radiusincrement 2 div def}%
\w@rps {elemcount \l@br startangle angleincrement add dup}%
\w@rps {cos startradius mul centerx add exch}%
\w@rps {sin startradius mul centery add}%
\w@rps {/startradius startradius radiusincrement add}%
\w@rps {excursion sub def}%
\w@rps {startangle angleincrement add dup}%
\w@rps {cos startradius mul centerx add exch}%
\w@rps {sin startradius mul centery add}%
\w@rps{startangle angleincrement smallincrement add 2 div add dup}%
\w@rps {cos startradius mul centerx add exch}%
\w@rps {sin startradius mul centery add}%
\w@rps {curveto}%
\w@rps {/startangle startangle angleincrement smallincrement add add def}%
\w@rps {startangle angleincrement sub dup}%
\w@rps {cos startradius mul centerx add exch}%
\w@rps {sin startradius mul centery add}%
\w@rps {/startradius startradius radiusincrement add}%
\w@rps {excursion add def}%
\w@rps {startangle angleincrement sub dup}%
\w@rps {cos startradius mul centerx add exch}%
\w@rps {sin startradius mul centery add}%
\w@rps {startangle dup}%
\w@rps {cos startradius mul centerx add exch}%
\w@rps {sin startradius mul centery add}%
\w@rps {curveto\r@br repeat}%
\w@rps {stroke grestore end}%
\w@rps {\r@br 0 dbdef}%
\w@rps {/blob \l@br}%
\w@rps {0 begin st gsave}%
\w@rps {dup type dup}%
\w@rps {/stringtype eq}%
\w@rps {\l@br pop 0 get }%
\w@rps {dup (B) 0 get eq dup 2 index}%
\w@rps {(L) 0 get eq or /ifleft exch def}%
\w@rps {exch (R) 0 get eq or /ifright exch def}%
\w@rps {/iffill false def \r@br}%
\w@rps {\l@br /ifleft false def}%
\w@rps {/ifright false def}%
\w@rps {/booleantype eq }%
\w@rps {{/iffill exch def}}%
\w@rps {{setgray /iffill true def} ifelse \r@br}%
\w@rps {ifelse}%
\w@rps {/freq exch def}%
\w@rps {/angle exch def}%
\w@rps {/yrad  exch def}%
\w@rps {/xrad  exch def}%
\w@rps {/y exch def}%
\w@rps {/x exch def}%
\w@rps {newpath}%
\w@rps {svm currentmatrix pop}%
\w@rps {x y translate   }%
\w@rps {angle rotate}%
\w@rps {xrad yrad scale}%
\w@rps {0 0 1 0 360 arc}%
\w@rps {gsave 1 setgray fill grestore}%
\w@rps {gsave svm setmatrix stroke grestore}%
\w@rps {gsave iffill {fill} if grestore}%
\w@rps {clip newpath}%
\w@rps {gsave }%
\w@rps {ifleft  \l@br -3 freq 3 { -1 moveto 2 2 rlineto} for}%
\w@rps {svm setmatrix stroke\r@br if }%
\w@rps {grestore}%
\w@rps {ifright \l@br 3 freq neg -3 { -1 moveto -2 2 rlineto} for}%
\w@rps {svm setmatrix stroke\r@br if}%
\w@rps {grestore end}%
\w@rps {\r@br 0 dbdef}%
\w@rps {/BSpl \l@br}%
\w@rps {0 begin}%
\w@rps {storexyn}%
\w@rps {currentpoint newpath moveto}%
\w@rps {n 1 gt \l@br}%
\w@rps {0 0 0 0 0 0 1 1 true subspline}%
\w@rps {n 2 gt \l@br}%
\w@rps {0 0 0 0 1 1 2 2 false subspline}%
\w@rps {1 1 n 3 sub \l@br}%
\w@rps {/i exch def}%
\w@rps {i 1 sub dup i dup i 1 add dup i 2 add dup false subspline}%
\w@rps {\r@br for}%
\w@rps {n 3 sub dup n 2 sub dup n 1 sub dup 2 copy false subspline}%
\w@rps {\r@br if}%
\w@rps {n 2 sub dup n 1 sub dup 2 copy 2 copy false subspline}%
\w@rps {\r@br if}%
\w@rps {end}%
\w@rps {\r@br 0 dbdef}%
\w@rps {/midpoint \l@br}%
\w@rps {0 begin}%
\w@rps {/y1 exch def}%
\w@rps {/x1 exch def}%
\w@rps {/y0 exch def}%
\w@rps {/x0 exch def}%
\w@rps {x0 x1 add 2 div}%
\w@rps {y0 y1 add 2 div}%
\w@rps {end}%
\w@rps {\r@br 0 dbdef}%
\w@rps {/thirdpoint \l@br}%
\w@rps {0 begin}%
\w@rps {/y1 exch def}%
\w@rps {/x1 exch def}%
\w@rps {/y0 exch def}%
\w@rps {/x0 exch def}%
\w@rps {x0 2 mul x1 add 3 div}%
\w@rps {y0 2 mul y1 add 3 div}%
\w@rps {end}%
\w@rps {\r@br 0 dbdef}%
\w@rps {/subspline \l@br}%
\w@rps {0 begin}%
\w@rps {/movetoNeeded exch def}%
\w@rps {y exch get /y3 exch def}%
\w@rps {x exch get /x3 exch def}%
\w@rps {y exch get /y2 exch def}%
\w@rps {x exch get /x2 exch def}%
\w@rps {y exch get /y1 exch def}%
\w@rps {x exch get /x1 exch def}%
\w@rps {y exch get /y0 exch def}%
\w@rps {x exch get /x0 exch def}%
\w@rps {x1 y1 x2 y2 thirdpoint}%
\w@rps {/p1y exch def}%
\w@rps {/p1x exch def}%
\w@rps {x2 y2 x1 y1 thirdpoint}%
\w@rps {/p2y exch def}%
\w@rps {/p2x exch def}%
\w@rps {x1 y1 x0 y0 thirdpoint}%
\w@rps {p1x p1y midpoint}%
\w@rps {/p0y exch def}%
\w@rps {/p0x exch def}%
\w@rps {x2 y2 x3 y3 thirdpoint}%
\w@rps {p2x p2y midpoint}%
\w@rps {/p3y exch def}%
\w@rps {/p3x exch def}%
\w@rps {movetoNeeded \l@br p0x p0y moveto \r@br if}%
\w@rps {p1x p1y p2x p2y p3x p3y curveto}%
\w@rps {end}%
\w@rps {\r@br 0 dbdef}%
\w@rps {/storexyn \l@br}%
\w@rps {0 begin}%
\w@rps {/n exch def}%
\w@rps {/y n array def}%
\w@rps {/x n array def}%
\w@rps {n 1 sub -1 0 \l@br}%
\w@rps {/i exch def}%
\w@rps {y i 3 2 roll put}%
\w@rps {x i 3 2 roll put}%
\w@rps {\r@br for end}%
\w@rps {\r@br 0 dbdef}%
\w@rps {/bop \l@br save 0 begin /saved exch def end}%
\w@rps {scale setlinecap setlinejoin setlinewidth setdash moveto}%
\w@rps {\r@br 1 dbdef}%
\w@rps {/eop {stroke 0 /saved get restore showpage} 1 dbdef}%
\w@rps {end /defineresource where}%
\w@rps {{pop mark exch /TeXDraw exch /ProcSet defineresource cleartomark}}%
\w@rps {{/TeXDraw exch readonly def} ifelse}%
\w@rps {/setpacking where {pop setpacking} if}%
\w@rps {/setglobal where {pop setglobal} if}%
\w@rps {\p@p EndResource}%
\w@rps {\p@p EndProlog}%
\w@rps {\p@p Page: 1 1}%
\w@rps {\p@p PageBoundingBox: (atend)}%
\w@rps {\p@p BeginPageSetup}%
\w@rps {/TeXDraw /findresource where}%
\w@rps {{pop /ProcSet findresource}}%
\w@rps {{load} ifelse}%
\w@rps {begin}%
\w@rps {0 0 [] 0 3 1 1 \p@sfactor\space \p@sfactor\space bop}%
\w@rps {\p@p EndPageSetup}%
}
\def\t@drclose {%
\pixtobp \xminpix \l@lxbp  \pixtobp \yminpix \l@lybp
\pixtobp \xmaxpix \u@rxbp  \pixtobp \ymaxpix \u@rybp
\w@rps {\p@p PageTrailer}%
\w@rps {\p@p PageBoundingBox: \the\l@lxbp\space \the\l@lybp\space
\the\u@rxbp\space \the\u@rybp}%
\w@rps {eop end}%
\w@rps {\p@p Trailer}%
\w@rps {\p@p BoundingBox: \the\l@lxbp\space \the\l@lybp\space
\the\u@rxbp\space \the\u@rybp}%
\w@rps {\p@p EOF}%
\closeout\drawfile}
\catcode`\@=\catamp
\def\dvialwsetup{
\def\includepsfile##1##2##3##4##5{\special{Insert ##1\space%
}}%
\def\rotsclTeX##1##2##3##4{\special{Insert /dev/null do %
3 index exch translate cleartomark %
matrix currentmatrix aload pop %
7 6 roll restore matrix astore %
matrix currentmatrix exch setmatrix %
0 0 moveto setmatrix %
gsave currentpoint 2 copy translate ##1 rotate %
##2 ##3 scale neg exch neg exch translate %
save}%
##4%
\special{Insert /dev/null do cleartomark restore %
currentpoint grestore moveto save}}%
}
\def\dvipssetup{
\def\includepsfile##1##2##3##4##5{\vbox to 0pt{%
\vskip##5%
\includegraphics{##1}%
\vss}}
\def\rotsclTeX##1##2##3##4{%
##4
}}
\dvipssetup


\chardef\catamp=\the\catcode`\@
\catcode`\@=11
\def\realadd #1#2#3{\dimen0=#1pt
\dimen2=#2pt
\advance \dimen0 by \dimen2
\edef #3{\expandafter\c@lean\the\dimen0}}
\def\realdiv #1#2#3{\dimen0=#1pt
\t@counta=\dimen0
\dimen0=#2pt
\t@countb=\dimen0
\intdiv \t@counta \t@countb #3}
\def\lenhyp #1#2#3{\t@counta=#1%
\multiply \t@counta by \t@counta
\t@countb=#2%
\multiply \t@countb by \t@countb
\advance \t@counta by \t@countb
\sqrtnum \t@counta #3}
\let\bk=\t@counta
\let\bn=\t@countb
\let\xval=\t@countc
\def\sqrtnum #1#2{\xval=#1%
\bk=\xval
\loop
\bn=\xval
\divide \bn by \bk
\advance \bn by \bk
\advance \bn by 1
\divide \bn by 2
\ifnum \bn < \bk
\bk=\bn
\repeat
#2=\bn}
\def\currentpos #1#2{\t@pixa=\x@pix
\advance \t@pixa by -\x@segoffpix
\pixtocoord \t@pixa #1
\t@pixa=\y@pix
\advance \t@pixa by -\y@segoffpix
\pixtocoord \t@pixa #2}
\def\vectlen (#1 #2)(#3 #4)#5{\getpos (#1 #2)\x@arga\y@arga
\getpos (#3 #4)\x@argb\y@argb
\coordtopix \x@arga \t@pixa
\coordtopix \x@argb \t@pixb
\advance \t@pixb by -\t@pixa
\coordtopix \y@arga \t@pixc
\coordtopix \y@argb \t@pixd
\advance \t@pixd by -\t@pixc
\lenhyp \t@pixb \t@pixd \t@pixc
\pixtocoord \t@pixc #5}
\def\cossin (#1 #2)(#3 #4)#5#6{\getpos (#1 #2)\x@arga\y@arga
\getpos (#3 #4)\x@argb\y@argb
\coordtopix \x@arga \t@pixa
\coordtopix \x@argb \t@pixb
\advance \t@pixb by -\t@pixa
\coordtopix \y@arga \t@pixc
\coordtopix \y@argb \t@pixd
\advance \t@pixd by -\t@pixc
\lenhyp \t@pixb \t@pixd \t@pixc
\intdiv \t@pixb\t@pixc #5%
\intdiv \t@pixd\t@pixc #6}
\catcode`\@=\catamp

\expandafter\chardef\csname m@catamp\endcsname=\the\catcode`\@
\catcode`\@=11
\def\straightness #1 {\realdiv 1{#1}\m@factor}

\def\gvec (#1 #2) {\currentpos \m@xpos\m@ypos
\realadd {#1}\m@xpos\m@
\realdiv \m@2\m@xav
\realadd {#2}\m@ypos\m@
\realdiv \m@2\m@yav
\edef\m@{(\m@xpos\space\m@ypos)}
\expandafter\cossin \m@(#1 #2)\m@cos\m@sin
\realdiv \m@sin{-\m@factor}\m@
\realadd \m@\m@xav\m@xcntr
\realdiv \m@cos\m@factor\m@
\realadd \m@\m@yav\m@ycntr
\edef\m@{around (\m@xcntr\space\m@ycntr) %
from (\m@xpos\space\m@ypos) to (#1 #2) }
\expandafter\drawcurvygluon \m@}

\def\pvec (#1 #2) {\currentpos \m@xpos\m@ypos
\realadd {#1}\m@xpos\m@
\realdiv \m@2\m@xav
\realadd {#2}\m@ypos\m@
\realdiv \m@2\m@yav
\edef\m@{(\m@xpos\space\m@ypos)}
\expandafter\cossin \m@(#1 #2)\m@cos\m@sin
\realdiv \m@sin{-\m@factor}\m@
\realadd \m@\m@xav\m@xcntr
\realdiv \m@cos\m@factor\m@
\realadd \m@\m@yav\m@ycntr
\edef\m@{around (\m@xcntr\space\m@ycntr) %
from (\m@xpos\space\m@ypos) to (#1 #2) }
\expandafter\drawcurvyphoton \m@}

\straightness{10}
\catcode`\@=\m@catamp
\straightness 100 


\newpage
\thispagestyle{empty}

\newcommand{\CGglghVertex}{
\begin{texdraw} \drawdim cm \setunitscale 0.35 \curvylength {15}
\curvyheight {10} \arrowheadtype t:F \arrowheadsize l:0.8 w:0.4
\move (-6.1 -3) \lpatt(.7 .3) \lvec(0 0) \move(6.1 -3) \lvec(0 0) \lpatt()
\pvec(0 6) \move(-3.2 -1.6) \avec(-2.6 -1.3) \move(3.6 -1.8) \avec(4 -2)
\arrowheadsize l:0.9 w:0.45 
\move(3.1 -3.3)\htext{$P$}
\move(-2 4.8) \htext{$\mu$} \move(1 5.1) \htext{$b$}
\move(-6.3 -2.2) \htext{$c$} \move(5.4 -2.2) \htext{$a$}
\move(-10 -10)\htext{$g f^{abc}\Gamma_{\mu}(P)=g f^{abc}\Bigl[a(P)P_{\mu}
   +b(P)\widetilde n_{\mu}(P)\Bigr]$}
\end{texdraw}}

\vbox to \vsize{
\tabskip=0cmplus1fil
\halign to \hsize{&\hfil#\hfil\cr
\noalign{\vskip 0cmplus1fil}
   \CGglghVertex          &\cr
\noalign{\vskip 0cmplus1fil}
\noalign{\vskip 0cmplus1fil}
                &$\mbox{Figure~1}$       &\cr
}}

\newpage
\thispagestyle{empty}

\newcommand{\grSEB}{
\begin{texdraw} \drawdim cm \setunitscale 0.25 \curvylength {15}
\curvyheight {10} \arrowheadtype t:F \arrowheadsize l:0.6 w:0.5
\drawcurvyphoton around (0 0) from (-2.3 0) to (2.3 0)
\pvec (6 0)
\drawcurvyphoton around (0 0) from (2.3 0) to (-2.3 0)
\pvec (-6 0)
\grayblob {0.0}
\drawblob xsize:0.4 ysize:0.4 at (-2.3 0)
\drawblob xsize:0.4 ysize:0.4 at (2.3 0)
\drawblob xsize:0.4 ysize:0.4 at (0 2.3)
\drawblob xsize:0.4 ysize:0.4 at (0 -2.3)
\htext (-1 -6) {(a)}
\end{texdraw}}

\newcommand{\grSEC}{
\begin{texdraw} \drawdim cm \setunitscale 0.25 \curvylength {15}
\curvyheight {10} \arrowheadtype t:F \arrowheadsize l:0.6 w:0.3
\move (-6 -1) \pvec (0 -1) \move (0 2)
\drawcurvyphoton around (-0.25 1.25) from (1.25 2.25) to (1.25 2.25)
\move (0 -1) \pvec (6 -1)
\grayblob {0.0}
\drawblob xsize:0.4 ysize:0.4 at (-0.25 3.3)
\drawblob xsize:0.4 ysize:0.4 at (-0.25 -1)

\htext (-1 -6) {(b)}
\end{texdraw}}

\newcommand{\grSED}{
\begin{texdraw} \drawdim cm \setunitscale 0.25 \curvylength {15}
\curvyheight {10} \arrowheadtype t:F \arrowheadsize l:0.6 w:0.5
\move (-6 0) \pvec (-2.3 0) \move (0 0) \lpatt(.7 .3) \lcir r:2.3 \lpatt()
\move (2.3 0) \pvec (6 0) \move (0 2.3) \avec (0.5 2.3) \move (0 -2.3)
\avec (-0.5 -2.3)
\htext (-1 -6) {(c)}
\end{texdraw}}

\vbox to \vsize{
\tabskip=0cmplus1fil
\halign to \hsize{&\hfil#\hfil\cr
\noalign{\vskip 0cmplus1fil}
\noalign{\vskip 3cm}
        \grSEB  &\grSEC         &\grSED\cr
\noalign{\vskip 0cmplus1fil}
\noalign{\vskip 0cmplus1fil}
                &Figure~2       &\cr
}}

\newpage
\thispagestyle{empty}

\newcommand{\grSE}{
\begin{texdraw} \drawdim cm \setunitscale 0.29 \curvylength {15}
\curvyheight {10} \arrowheadtype t:F \arrowheadsize l:0.6 w:0.3
\move (-5 4) \lvec (-5 -4) \move (-5 2) \avec (-5 2.5) \move (-5 -2.5)
\avec (-5 -2) \move (5 4) \lvec (5 -4) \move (5 2) \avec (5 2.5)
\move (5 -2.5) \avec (5 -2) \move (-5 0) \pvec (0 0) \pvec (5 0)
\grayblob {0.8} \drawblob xsize:2.5 ysize:2.5 at (0 0) \move (0 0)
\lcir r:2.5 
\grayblob {0.0}
\drawblob xsize:0.4 ysize:0.4 at (-3.8 0)
\drawblob xsize:0.4 ysize:0.4 at (3.8 0)
\end{texdraw}}

\vbox to \vsize{
\tabskip=0cmplus1fil
\halign to \hsize{&\hfil#\hfil\cr
\noalign{\vskip 0cmplus1fil}
                &\grSE          &\cr
\noalign{\vskip 0cmplus1fil}
\noalign{\vskip 0cmplus1fil}
                &Figure~3       &\cr
}}

\newpage
\thispagestyle{empty}

\newcommand{\grVertexFirstone}{\begin{texdraw} \drawdim cm \setunitscale 0.25
\curvylength {15} \curvyheight {10} \arrowheadtype t:F 
\arrowheadsize l:0.6 w:0.3 \move (-5 4) \lvec (-5 -4) \move (5 4) \lvec (5 -4)
\drawcurvyphoton around (-5 0) from (-5 2.5) to (-5 -2.5)
\move (-5 0) \pvec (0 0) \pvec (5 0)
\grayblob {0.0}
\drawblob xsize:0.4 ysize:0.4 at (0 0)
\drawblob xsize:0.4 ysize:0.4 at (-7.5 0)
\htext (9 0) {+} 
\htext (9 -8) {(a)} 
\end{texdraw}}

\newcommand{\grVertexFirsttwo}{\begin{texdraw} \drawdim cm \setunitscale 0.25
\curvylength {15}
\curvyheight {10} \arrowheadtype t:F \arrowheadsize l:0.6 w:0.3
\move (-5 4) \lvec (-5 -4) \move (5 4) \lvec (5 -4)
\drawcurvyphoton around (-5 0.75) from (-5 2.75) to (-5 -1.25)
\move (-5 -2.5) \pvec (0 -2.5) \pvec (5 -2.5)
\grayblob {0.0}
\drawblob xsize:0.4 ysize:0.4 at (0 -2.5)
\drawblob xsize:0.4 ysize:0.4 at (-7 0.75)
\htext (10 0) {$\Longrightarrow$} 
\htext (10 -8) {}
\end{texdraw}}

\newcommand{\grPinchVertexFirst}{\begin{texdraw} \drawdim cm \setunitscale 0.25
\curvylength {15} \curvyheight {10} \arrowheadtype t:F
\arrowheadsize l:0.6 w:0.3
\move (-1 4) \lvec (-1 -4) \move (9 4) \lvec (9 -4)
\drawcurvyphoton around (-3.3 0) from (-3.3 -2.1) to (-3.3 -2.1)
\move (-1 0) \pvec (4 0) \pvec (9 0)
\grayblob {0.0}
\drawblob xsize:0.4 ysize:0.4 at (4 0)
\drawblob xsize:0.4 ysize:0.4 at (-5.7 0)
\htext (4 -8) {(b)} \htext (-7 0) {}
\end{texdraw}}



\newcommand{\grVertexSecond}{\begin{texdraw} \drawdim cm \setunitscale 0.25 
\curvylength {15}
\curvyheight {10} \arrowheadtype t:F \arrowheadsize l:0.6 w:0.3
\move (-5 4) \lvec (-5 -4) \move (5 4) \lvec (5 -4)
\move (-5 2.3) \pvec (-1.6 0) \pvec (2.5 0) \pvec (5 0) \move (-1.6 0)
\pvec (-5 -2.3)
\grayblob {0.0}
\drawblob xsize:0.4 ysize:0.4 at (-1.6 0)
\drawblob xsize:0.4 ysize:0.4 at (2.5 0)
\drawblob xsize:0.4 ysize:0.4 at (-3.3 1.15)
\drawblob xsize:0.4 ysize:0.4 at (-3.3 -1.15)
\htext (0 -8) {(a)} \htext (20 0) {$\Longrightarrow$} 
\end{texdraw}}


\newcommand{\grPinchVertexSecond}{\begin{texdraw} \drawdim cm 
\setunitscale 0.25 \curvylength {15} \curvyheight {10} \arrowheadtype t:F 
\arrowheadsize l:0.6 w:0.3
\move (-5 4) \lvec (-5 -4) \move (5 4) \lvec (5 -4)
\drawcurvyphoton around (-2.5 1.25) from (-5 0) to (0 0)
\drawcurvyphoton around (-2.5 -1.25) from (0 0) to (-5 0)
\move (0 0) \pvec (5 0) 
\grayblob {0.0}
\drawblob xsize:0.4 ysize:0.4 at (2.5 0)
\drawblob xsize:0.4 ysize:0.4 at (0 0)
\drawblob xsize:0.4 ysize:0.4 at (-2.5 1.55)
\drawblob xsize:0.4 ysize:0.4 at (-2.5 -1.55)
\htext (0 -8) {(b)}
\end{texdraw}}


\newcommand{\grBoxone}{\begin{texdraw} \drawdim cm \setunitscale 0.25 
\curvylength {15}
\curvyheight {10} \arrowheadtype t:F \arrowheadsize l:0.6 w:0.3
\move (-5 4) \lvec (-5 -4) \move (5 4) \lvec (5 -4)
\move (-5 2) \pvec (0 2) \pvec (5 2) \move (-5 -2) \pvec (0 -2) \pvec (5 -2)
\grayblob {0.0}
\drawblob xsize:0.4 ysize:0.4 at (0 2)
\drawblob xsize:0.4 ysize:0.4 at (0 -2)
\htext (10 0) {+}
\htext (10 -8) {(a)}
\end{texdraw}}

\newcommand{\grBoxtwo}{\begin{texdraw} \drawdim cm \setunitscale 0.25 
\curvylength {15}
\curvyheight {10} \arrowheadtype t:F \arrowheadsize l:0.6 w:0.3
\move (-5 4) \lvec (-5 -4) \move (5 4) \lvec (5 -4) \move (-5 2)
\pvec (0 0) \pvec (5 -2) \move (5 2) \pvec (0 0) \pvec (-5 -2)
\grayblob {0.0}
\drawblob xsize:0.4 ysize:0.4 at (2.5 1)
\drawblob xsize:0.4 ysize:0.4 at (2.5 -1)
\htext (13 0) {$\Longrightarrow$}
\htext (0 -8) {}
\end{texdraw}}

\newcommand{\grBoxP}{ \begin{texdraw} \drawdim cm \setunitscale 0.25 
\curvylength {15}
\curvyheight {10} \arrowheadtype t:F \arrowheadsize l:0.6 w:0.3
\move (-5 4) \lvec (-5 -4) \move (5 4) \lvec (5 -4)
\drawcurvyphoton around (0 -3) from (5 0) to (-5 0)
\drawcurvyphoton around (0 3) from (-5 0) to (5 0)
\grayblob {0.0}
\drawblob xsize:0.4 ysize:0.4 at (0 2.8)
\drawblob xsize:0.4 ysize:0.4 at (0 -2.8)
\htext (0 -8) {(b)}
\end{texdraw}}


\vbox to \vsize{
\tabskip=0cmplus1fil
\halign to \hsize{#&#&#\cr
\noalign{\vskip 0cmplus1fil}
\grVertexFirstone & \hspace{-1.5cm}\grVertexFirsttwo    &\hspace{1cm}
                                                 \grPinchVertexFirst \cr
\noalign{\vskip 0cmplus1fil}
                        &$\mbox{Figure~4}$              &\cr
\noalign{\vskip 0cmplus1fil}
        \qquad\quad\grVertexSecond &            &\qquad\grPinchVertexSecond \cr
\noalign{\vskip 0cmplus1fil}
                        &$\mbox{Figure~5}$              &\cr
\noalign{\vskip 0cmplus1fil}
        \grBoxone       & \hspace{-1.5cm}\grBoxtwo &\quad\grBoxP \cr
\noalign{\vskip 0cmplus1fil}
                        &$\mbox{Figure~6}$              &\cr
\noalign{\vskip 0cmplus1fil}
}}

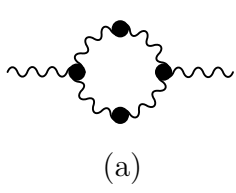
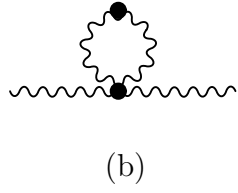
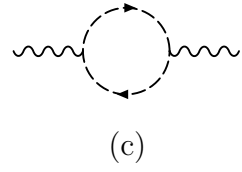
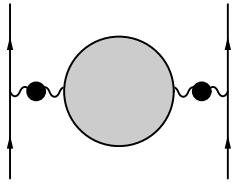
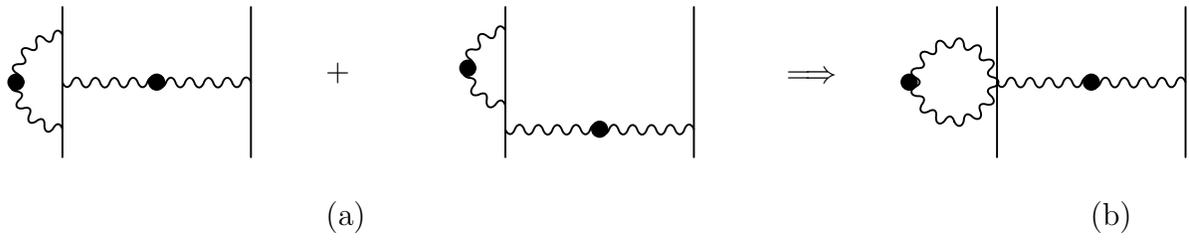
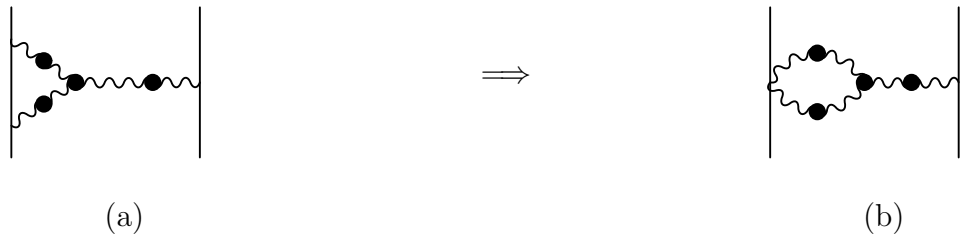
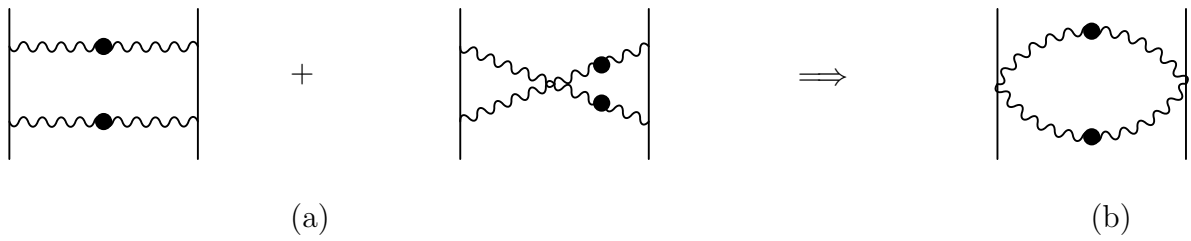
\end{document}